\documentclass[aps,prx,reprint, amsmath, amssymb,superscriptaddress,longbibliography]{revtex4-2}

\usepackage{bm}
\usepackage[retainorgcmds]{IEEEtrantools}
\usepackage{graphicx}
\usepackage{mathrsfs}
\usepackage{color}
\usepackage{times,txfonts}
\usepackage{nicefrac}
\usepackage{ragged2e}
\usepackage{tikz}
\usepackage{physics}

\usepackage[colorlinks=true,linkcolor=blue,urlcolor=blue,citecolor=blue,pdfusetitle]{hyperref}

\usepackage{blindtext}

\newcommand{\e}{\mathrm{e}}
\newcommand{\diff}{\mathrm{d}}
\newcommand{\I}{\mathrm{i}}

\newcommand\mydots{\hbox to 1em{.\hss.\hss.}}

\newcommand{\sgn}{{\rm sgn}}

\begin{document}

\title{Dissipative framework for subsystem dynamics of noninteracting quantum chains}

\author{Michele Coppola}
\affiliation{Jožef Stefan Institute, SI-1000 Ljubljana, Slovenia}
\author{Mari Carmen Ba\~nuls}
\affiliation{Max-Planck-Institut für Quantenoptik, Hans-Kopfermann-Str. 1, D-85748 Garching, Germany}
\affiliation{Munich Center for Quantum Science and Technology (MCQST), Schellingstr. 4, D-80799 München, Germany}
\author{Zala Lenar\v{c}i\v{c}}
\affiliation{Jožef Stefan Institute, SI-1000 Ljubljana, Slovenia}

\begin{abstract}
When only local observables of a many-body quantum system are of interest, it is desirable to formulate a reduced description within the Hilbert space of the corresponding subsystem, with the remaining degrees of freedom traced out and acting as an environment. Assuming initially uncorrelated states and Gaussian environments, we develop a framework for reconstructing the local dynamical generator of noninteracting quantum chains, with polynomial computational effort. As an application, we consider two representative models: a bipartitioned Kitaev chain and a Kitaev chain boundary-coupled to a fully connected free-fermion environment. In both models, strong subsystem-environment coupling leads to non-Markovian dynamics characterized by ballistic spreading of the Lindblad dissipator support within the subsystem. On the other hand, weak coupling to a fully connected environment yields predominantly boundary-localized, Markovian dissipation. Our work highlights the implications of subsystem-environment correlations on the generator of local dynamics, beyond the conventional weak-coupling approximations. 
\end{abstract}

\maketitle{}

\section{Introduction}
Many-body quantum systems can always be partitioned into a subsystem and its complement, which acts as an external environment.
When the objective is to characterize only subsystem observables, solving the full many-body unitary dynamics of the composite state is often unnecessary. The theory of open quantum systems aims to derive a reduced description formulated entirely within the Hilbert space of the subsystem. This is achieved by constructing a generator of the local dynamics that effectively incorporates the influence of the external environment on the subsystem. In the weak-coupling limit between the subsystem and the environment, and assuming the Born-Markov and secular approximations, the reduced dynamics is described by the celebrated Lindblad master equation~\cite{breuer2002theory}. However, deriving the generator of the subsystem dynamics beyond these assumptions remains one of the central challenges of open quantum systems. Prominent approaches include the Schwinger-Keldysh contour formalism~\cite{schwinger1961brownian,keldysh2024diagram,kamenev2023field,reimer2019density,d2025exact}, the process tensor framework~\cite{strathearn2018efficient,cygorek2022simulation,gribben2022exact,fowler2022efficient,fux2023tensor}, and more recently machine-learning methods for reconstructing time-local generators~\cite{mazza2021machine,cemin2024inferring,coppola2025learning}.

Beyond the traditional open-system setting, reduced subsystem dynamics has also emerged as a powerful perspective for understanding the nonequilibrium dynamics of isolated many-body systems. Although global evolution is unitary, finite subsystems generally undergo effective irreversible dynamics as they become entangled with the rest of the system. This viewpoint has become increasingly important in the study of local equilibration, thermalization, information spreading, and hydrodynamic behavior. Recent developments have introduced efficient representations of subsystem evolution, including the influence matrix formalism~\cite{strathearn2018efficient,lerose2023overcoming,lerose2021influence,sonner2021influence,guo2024efficient}, transverse light-cone contractions~\cite{perez24}, information lattice methods~\cite{kvorning2022local,artiaco24,artiaco24a}, and related tensor-network approaches for subsystem propagators~\cite{rakovszky2022daoe}. These methods suggest that the reduced dynamics of a subsystem contains information about the emergence of irreversibility and memory effects in isolated quantum matter, even in the absence of an external bath.

Motivated by these complementary perspectives, we investigate the subsystem dynamics in one-dimensional lattices of non-interacting quantum particles, considering both fermionic and bosonic models. Such systems provide an ideal setting in which the environment is fully microscopic while remaining computationally tractable, allowing the interplay between subsystem-environment correlations, non-Markovian effects, and effective dissipation to be analyzed exactly. Assuming initially uncorrelated subsystem-environment states and Gaussian environments, we develop a framework that reconstructs the exact time-local dynamical generator with polynomial computational effort, avoiding the exponential scaling typically associated with the subsystem Hilbert space. The resulting Liouvillian enables a direct characterization of the emerging Lindblad operators, their time dependence, and the degree of non-Markovianity induced by the strong coupling to the environment.

\begin{figure}
    \centering
    \includegraphics[width=\columnwidth]{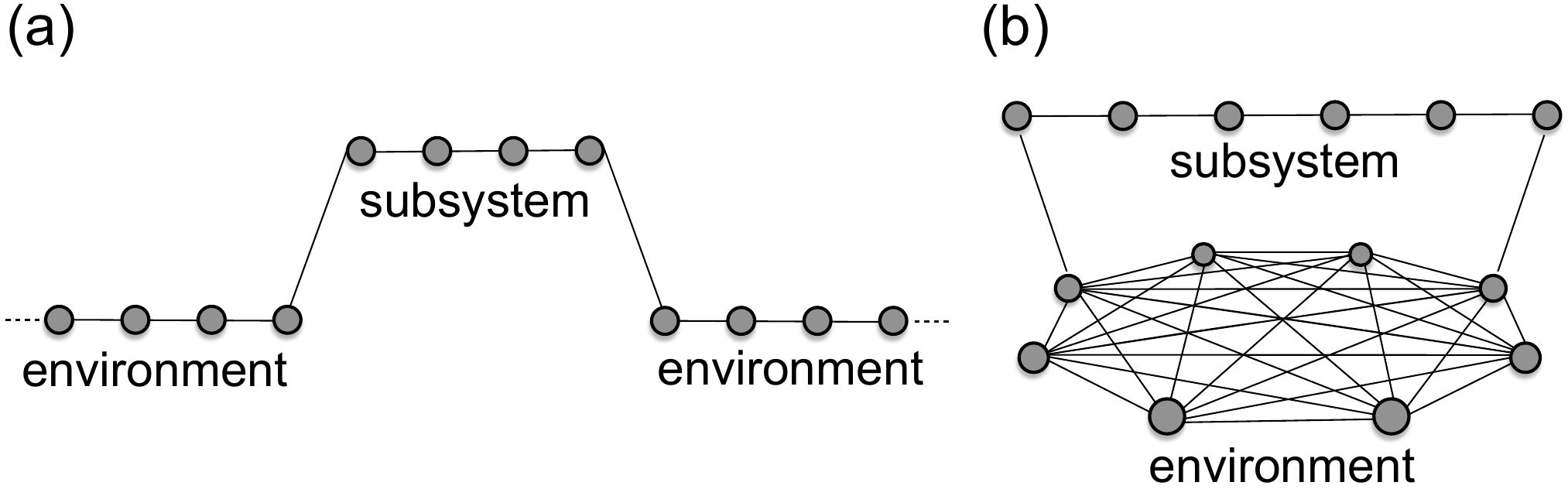}
    \caption{Schematic representations of the models studied in this work: (a) a bipartitioned Kitaev chain; (b) a Kitaev chain boundary-coupled to a fully connected environment.}
    \label{fig1}
\end{figure}

As concrete applications, we study two models (see Fig.~\ref{fig1}): (a) a Kitaev chain partitioned into subsystem and environment and (b) a Kitaev chain boundary-coupled to a fully connected free-fermion environment. We examine how the generator of the local dynamics depends on the subsystem size, model parameters, environment structure, and subsystem-environment coupling.
Beyond providing an exact benchmark for open-system descriptions outside the weak-coupling regime, these models are used to establish a direct connection between microscopic many-body dynamics and emergent reduced descriptions relevant to subsystem equilibration, information propagation, and thermalization in isolated quantum systems.

The remainder of this paper is organized as follows. In Secs.~\ref{Sec.II} and \ref{Sec.III}, we introduce the general framework, define the subsystem-environment setting, and derive the general form of the reduced time-local generator, under the assumptions of initially uncorrelated states and Gaussian environments. In Sec.~\ref{Sec.IV}, we present an efficient reconstruction of the time-dependent Liouvillian from the Lyapunov equation for the correlation matrix. In Secs.~\ref{Sec.V} and ~\ref{Sec.VI}, we apply the formalism to a Kitaev chain and a fully connected free-fermion environment, respectively. We characterize the emergence of Markovian or non-Markovian dynamics, the structure of the Lindblad operators, and their dependence on the environment temperature, subsystem size, and model parameters. Finally, Sec.~\ref{Sec.VII} summarizes our main findings and discusses future directions. Technical derivations are collected in the Appendixes.

%%%%%%%%%%%%%%%%%%%%%%%%%%%%%%%%%%%%%%%%%%%
%%%%%%%%%%%%%%%%%%%%%%%%%%%%%%%%%%%%%%%%%%%

\section{General framework}
\label{Sec.II}

Let us consider a closed many-body quantum system, which we partition into a \emph{subsystem} $(\rm S)$ and its complement, known as the external \emph{environment} $(\rm E)$ in the literature on open quantum systems. The total Hilbert space is $\mathcal{H}=\mathcal{H}^{\rm S}\otimes \mathcal{H}^{\rm E}$, where $\mathcal{H}^{\rm S}$ and $\mathcal{H}^{\rm E}$ denote the Hilbert spaces of the subsystem and the environment, respectively. The full subsystem-environment evolution is generated by the Hamiltonian
\begin{equation}
    H=H^{\rm S}\otimes \bm{\mathbb{I}}^{\rm E}+\bm{\mathbb{I}}^{\rm S}\otimes H^{\rm E} + H^{\rm I}\,,
\end{equation}
where $H^{\rm S}$ and $H^{\rm E}$ are the local Hamiltonians and $H^{\rm I}$ is the interaction Hamiltonian. The composite subsystem-environment state $\rho_t$ at an arbitrary time $t$ can always be decomposed as
\begin{equation}
    \rho_t =\rho^{\rm S}_t\otimes\rho^{\rm E}_t+\chi_t\,,
\end{equation}
where $\rho^{\rm S}_t=\tr_{\rm E}(\rho_t)$ and $\rho^{\rm E}_t=\tr_{\rm S}(\rho_t)$ denote the reduced density operators; $\chi_t$ is the traceless operator encoding all subsystem-environment correlations. Since the full many-body system is closed, the evolution of $\rho_t$ is governed by the Liouville equation 
\begin{equation}\label{liouville}
    \dot\rho_t=-\I \comm{H}{\rho_t}\,.
\end{equation}
Therefore, in the absence of initial subsystem-environment correlations ($\chi_0=0$), the reduced state reads
\begin{equation}\label{dynamical_map}
    \rho^{\rm S}_t := \epsilon_t\left[\rho^{\rm S}_0\right] =\tr_{\rm E}\left(\e^{-\I H t}\,\rho^{\rm S}_0\otimes\rho^{\rm E}_0\,\e^{\I H t}\right)\,.
\end{equation}
The linear map $\epsilon_t$ is the reduced dynamical map that provides the evolution of the subsystem state $\rho^{\rm S}_t$. 
We emphasise that $\epsilon_t$ is a Kraus-type map~\cite{breuer2002theory}, which is completely positive and trace-preserving (CPTP), i.e. $\epsilon_t$ is a proper \emph{quantum channel}. By definition, the quantum map $\epsilon_t$ is completely positive (CP) if the extended map $\epsilon_t\otimes \bm{\mathbb{I}}_{n}$, acting on the subsystem and any $n$-dimensional ancilla, is positive for all $n$.

Throughout this work, CP-divisibility will be used as a probe of Markovianity. We recall that the dynamical map $\epsilon_{t}$ is said to be divisible if there exists a dynamical propagator $\epsilon_{t,s}$ such that
\begin{equation}\label{divisibility}
    \epsilon_{t}=\epsilon_{t,s}\circ \epsilon_{s}\,,\qquad \forall s,t\geq 0\,.
\end{equation}
Notice that while $\epsilon_{t}$ and $\epsilon_{s}$ are CPTP maps by construction, $\epsilon_{t,s}$ need not be CP and not even
positive. Therefore, the dynamical map $\epsilon_{t}$ is said to be CP-divisible if $\epsilon_{t,s}$ exists and is CP $\forall s,t$. 
The composition law in Eq.~\eqref{divisibility} is the quantum counterpart of the Chapman-Kolmogorov equation for classical Markov processes, which expresses transition probabilities in terms of a sum over all possible intermediate configurations~\cite{breuer2016colloquium}. By analogy, it has been proposed to identify quantum Markovianity with CP-divisible evolutions~\cite{breuer2016colloquium}.  

The study of CP-divisibility proceeds through the analysis of the local generator of the subsystem dynamics.
By taking the first derivative of Eq.~\eqref{dynamical_map} and assuming that the dynamical map $\epsilon_t$ is not singular, one obtains
\begin{equation}
    \dot\rho^{\rm S}_t = \dot{\epsilon}_t \circ \epsilon_t^{-1}\left[\rho^{\rm S}_t\right]\,.
\end{equation}
Therefore one can define $\mathcal{L}_t := \dot{\epsilon}_t \circ\epsilon_t^{-1}$ as the generator of the local dynamics, namely the superoperator $\mathcal{L}_t$ such that $\dot{\rho}^{\rm S}_t=\mathcal{L}_t [{\rho}^{\rm S}_t]$. 
As shown in Ref.~\cite{hall2014canonical}, $\mathcal{L}_t$ can be expressed in the pseudo-Lindblad form
\begin{equation}\label{general_Lind}
\hspace{-0.2cm}\mathcal{L}_t [{\rho}^{\rm S}_t]=  \I\comm{\rho^{\rm S}_t}{H^{\rm eff}_t} +\sum_{i=1}^{\mathcal{D}_{\rm S}^2-1}\Gamma_i(t) \left(L^i_t \, {\rho}^{\rm S}_t L^{i\,\dag}_t- \frac{1}{2}\{L_t^{i\,\dag} L^i_t,{\rho}^{\rm S}_t \} \right)\,,  
\end{equation}
with time-dependent effective Hamiltonian $H^{\rm eff}_t$, couplings $\Gamma_i(t)$ and Lindblad operators $L^i_t$. In the most general scenario, the dissipator contains $\mathcal{D}_{\rm S}^2-1$ terms, where $\mathcal{D}_{\rm S}=\dim(\mathcal{H}^{\rm S})$ is the dimension of the Hilbert space $\mathcal{H}^{\rm S}$. Provided that the inverse dynamical map exists, Eq.~\eqref{general_Lind} leads to two main considerations: (i) the dynamical map is CP-divisible if and only if $\Gamma_i(t)\geq 0$~\cite{breuer2016colloquium}; (ii) the number of terms in the Liouvillian can scale up to quadratically with the dimension of the Hilbert space. For instance, for spinless fermions, this scaling is exponential in the subsystem size, making the problem practically intractable for extensive subsystems. 

The main goal of this work is to reconstruct and study the generator $\mathcal{L}_t$ of the local dynamics and examine its nature (non-Markovian vs Markovian). This will be carried out for noninteracting Hamiltonians $H$, initially uncorrelated states ($\chi_0=0$), and initially Gaussian environments $\rho_0^{\rm E}$. These hypotheses allow us to reconstruct the time-local generator with polynomial computational effort. 

%%%%%%%%%%%%%%%%%%%%%%%%%%%%%%%%%%%%%%%%%%%
%%%%%%%%%%%%%%%%%%%%%%%%%%%%%%%%%%%%%%%%%%%

\section{Local dynamical generator}
\label{Sec.III}

Let us consider a $L$-site lattice of noninteracting quantum particles, whose Hamiltonian reads 
\begin{equation}\label{Hamiltonian}
    H = \sum_{ij=1}^{L} \left[K_{ij} c_i^\dag c_{j} + \frac{1}{2}\left(G_{ij}c_i^\dag c^\dag_{j} + \text{h.c.}\right)\right]\,,
\end{equation}
where the $c$'s denote either the fermionic or bosonic creation and annihilation operators; the $L\times L$ coefficient matrices $\bm{K}$ and $\bm{G}$ satisfy $\bm{K}=\bm{K}^\dag$ and $\bm{G}^{\rm T}=\mp\bm{G}$, where the $-$ sign corresponds to fermions and the $+$ sign to bosons. Since both fermionic (F) and bosonic (B) cases can be treated with minimal modifications, for the purposes of this section only, the upper sign will refer to fermions and the lower to bosons. It is convenient to introduce the $2L$ operators, 
\begin{equation}\label{majorana_fermions}
y_{2i-1}=c^\dag_i + c_i\,,\qquad y_{2i}=\I (c^\dag_i - c_i)\,, \qquad\forall i\in [1,L]\,,
\end{equation}
which correspond, respectively, to Majorana operators for fermions and quadratures for bosons. The operators $y_i$ are Hermitian and satisfy the algebra, 
\begin{align}
\label{algebraF}\{y_i,y_j\}&=2\delta_{ij}\,, & \text{(F)}\\
\label{algebraB}[y_i,y_j]&=2\I \Omega_{ij}\,, & \text{(B)}
\end{align} 
$\forall i,j\in [1,2L]$, where $\bm{\Omega}=\I\bm{\mathbb{I}}_L\otimes\bm{\sigma}_y$ and $\bm{\mathbb{I}}_L$ is the $L\times L$ identity matrix. In this notation, up to additive terms proportional to the identity operator, the Hamiltonian~\eqref{Hamiltonian} can be rewritten as
\begin{equation}\label{Hamiltonian2}
    H = \frac{1}{4}\bm{y}^\dag \cdot \bm{T} \cdot\bm{y}\,,
\end{equation}
where $\bm{y}^\dag=(y_1,y_2,\dots,y_{2L})$ and $\bm{T}=\bm{T}^\dag=\mp \bm{T}^{\rm T}$ is a $2L\times 2L$ coefficient matrix.
See App.~\ref{App_Tmatrix} for details.

The unitary evolution generated by the Hamiltonian~\eqref{Hamiltonian2} is Gaussian-preserving, meaning that any normalized initial density matrix of the form $\rho_0=\e^{\frac{1}{4}\bm{y}^\dag \cdot \bm{M}_0 \cdot\bm{y}}/\tr(\e^{\frac{1}{4}\bm{y}^\dag \cdot \bm{M}_0 \cdot\bm{y}})$, with $\bm{M}_0=\bm{M}_0^\dag = \mp \bm{M}_0^{\rm T}$, evolves into
\begin{equation}
    \rho_t=\e^{-\I H t}\,
\rho_0\,\e^{\I H t}=\frac{\e^{\frac{1}{4}\bm{y}^\dag \cdot \bm{M}_t \cdot\bm{y}}}{\tr(\e^{\frac{1}{4}\bm{y}^\dag \cdot \bm{M}_t \cdot\bm{y}})}\,,
\end{equation}
with $\bm{M}_t=\bm{M}_t^\dag = \mp \bm{M}_t^{\rm T}$. Importantly, the dynamics is fully determined by the correlation matrix $\bm{\Theta}_t$, with elements
\begin{equation}
    {\Theta}_{ij}(t)=\begin{cases}
     \frac{1}{2}\tr\left([y_i,y_j]\rho_t\right)=\frac{1}{2}\langle [y_i,y_j]\rangle_t\qquad\text{(F)}\\
     \frac{1}{2}\tr\left(\{y_i,y_j\}\rho_t\right)=\frac{1}{2}\langle \{y_i,y_j\}\rangle_t\qquad\text{(B)}
    \end{cases}\,,
\end{equation}
and $\bm{\Theta}_t=\bm{\Theta}_t^\dag=\mp \bm{\Theta}_t^{\rm T}$.

We now partition the quantum chain into two segments: an $l$-site quantum chain, from site $1$ to site $l<L$, corresponding to the subsystem of interest, and an $(L-l)$-site chain, from site $l+1$ to site $L$, which we identify as the external environment. We further assume that $\rho^{\rm E}_0$ is a Gaussian state of the form $\rho^{\rm E}_0\propto\e^{\frac{1}{4}\bm{y}^\dag \cdot \bm{M}^{\rm E}_0 \cdot\bm{y}}$. Since we take $\rho_0=\rho^{\rm S}_0\otimes\rho^{\rm E}_0$, if $\rho^{\rm S}_0$ is also Gaussian, then both $\rho_0$ and $\rho_t$ are Gaussian states because the dynamics generated by the total subsystem-environment Hamiltonian~\eqref{Hamiltonian2} preserves Gaussianity. Consequently, $\rho^{\rm S}_t$ remains Gaussian as the partial trace of a Gaussian state is itself Gaussian. 

If we assume a time local form for the subsystem dynamics, Gaussianity constrains the local generator to take the form
\begin{equation}\label{Lindblad_ansatz}
    \mathcal{L}_t\left[\rho^{\rm S}_t\right] = \I\comm{\rho^{\rm S}_t}{H^{\rm eff}_t} + \sum_{ij=1}^{2l}{\gamma}_{ij}(t)\left[y_i\, \rho^{\rm S}_t y_j - \frac{1}{2}\{y_jy_i,\rho^{\rm S}_t\}\right]\,,
\end{equation}
where $\gamma_{ij}(t)$ are matrix elements of a $2l\times 2l$ Hermitian matrix $\bm{\gamma}_t$ and $H^{\rm eff}_t$ is an effective time-dependent Hamiltonian
\begin{equation}
    H^{\rm eff}_t=\frac{1}{4}\sum_{ij=1}^{2l} h^{\rm eff}_{ij}(t)\,y_i y_j\,,    
\end{equation}
with $h^{\rm eff}_{ij}(t)$ the elements of the $2l\times 2l$ matrix $\bm{h}^{\rm eff}_t=(\bm{h}^{\rm eff}_t)^\dag=\mp (\bm{h}^{\rm eff}_t)^{\rm T}$. The Liouvillian~\eqref{Lindblad_ansatz} represents the most general time-local and Gaussian-preserving generator of the subsystem dynamics. Importantly, the matrix $\bm{\gamma}_t$ is not required to be positive semidefinite, allowing for general non-Markovian evolutions, and the number of its eigenvalues, $\Gamma_i(t)$, scales linearly with the subsystem size $l$. 

%%%%%%%%%%%%%%%%%%%%%%%%%%%%%%%%%%%%%%%%%%%
%%%%%%%%%%%%%%%%%%%%%%%%%%%%%%%%%%%%%%%%%%%

\section{Deriving the Liouvillian}
\label{Sec.IV}

In order to reconstruct $\bm{h}^{\rm eff}_t$ and $\bm{\gamma}_t$, we start from the equation of motion for the correlation matrix $\bm{\Theta}_t$ of the entire $L$-site lattice. From the Liouville equation~\eqref{liouville}, it follows that
\begin{equation}\label{liouville2}
\dot{\bm{\Theta}}_t = 
\begin{cases}
 -\I\comm{\bm{T}}{\bm{\Theta}_t}\hspace{0.5cm} & \text{(F)}\\ 
 -\left(\bm{T}\bm{\Omega}\right)^{\rm T}\bm{\Theta}_t - \bm{\Theta}_t\left(\bm{T}\bm{\Omega}\right)\hspace{0.5cm}&\text{(B)}
\end{cases}\,,    
\end{equation}
which admits the solution
\begin{equation}\label{ev_corr_matrix}
\bm{\Theta}_t = 
\begin{cases}
\e^{-\I t \bm{T}}\,\bm{\Theta}_0\,\e^{\I t \bm{T}} \quad &   \text{(F)}\\ 
\e^{- t (\bm{T}\bm{\Omega})^{\rm T}}\,\bm{\Theta}_0\,\e^{-t (\bm{T}\bm{\Omega})}\quad & \text{(B)}\\   
\end{cases}\,.
\end{equation}
We decompose the matrices $\e^{-\I t \bm{T}}$, $\e^{- t (\bm{T}\bm{\Omega})^{\rm T}}$ and $\bm{\Theta}_t$ into blocks corresponding to the subsystem and environmental degrees of freedom. We organise  $2L\times 2L$ matrices $\bm{\mathcal{M}}$ in the following way: 
\begin{equation}\label{decomp}
\bm{\mathcal{M}} =\begin{bmatrix}
    \bm{\mathcal{M}}^{\rm S} & \bm{\mathcal{M}}^{\rm SE} \\
    \bm{\mathcal{M}}^{\rm ES} & \bm{\mathcal{M}}^{\rm E}
\end{bmatrix}\,,    
\end{equation}
where $\bm{\mathcal{M}}^{\rm S}$ and $\bm{\mathcal{M}}^{\rm E}$ denote the subsystem and environment truncated forms, while $\bm{\mathcal{M}}^{\rm SE}$ and $\bm{\mathcal{M}}^{\rm ES}$ describe the cross-couplings between them. The respective block dimensions are $2l\times 2l$ for $\bm{\mathcal{M}}^{\rm S}$, $2l\times 2(L-l)$ for $\bm{\mathcal{M}}^{\rm SE}$, $2(L-l)\times 2l$ for $\bm{\mathcal{M}}^{\rm ES}$, and $2(L-l)\times 2(L-l)$ for $\bm{\mathcal{M}}^{\rm E}$. Within this notation,  $\bm{\Theta}_t^{\rm S}$ and $\bm{\Theta}_t^{\rm E}$ denote the subsystem and environment correlation matrices, respectively. Importantly, the off-diagonal blocks in $\bm{\Theta}_0$ are zero, since the subsystem and environment are initially uncorrelated.

Using Eq.~\eqref{ev_corr_matrix} and the decomposition \eqref{decomp}, one can show that the subsystem correlation matrix $\bm{\Theta}_t^{\rm S}$ reads  
\begin{equation}\label{sol_red_correl_matrix}
    \bm{\Theta}_t^{\rm S}=\bm{U}_t\,\bm{\Theta}_0^{\rm S}\,\bm{U}_t^\dag+\bm{V}_t\,,
\end{equation}
where 
\begin{align}
\label{U_t}\bm{U}_t&:=
\begin{cases}
[\e^{-\I t \bm{T}}]^{\rm S}&\text{(F)}\\
[\e^{-t (\bm{T}\bm{\Omega})^{\rm T} }]^{\rm S}&\text{(B)}
\end{cases}\,,\quad\\    
\label{V_t}\bm{V}_t&:=
\begin{cases}
[\e^{-\I t \bm{T}}]^{\rm SE}\,\bm{\Theta}_0^{\rm E}\,[\e^{\I t \bm{T}}]^{\rm ES}&\text{(F)}\\
[\e^{-t (\bm{T}\bm{\Omega})^{\rm T} }]^{\rm SE}\,\bm{\Theta}_0^{\rm E}\,[\e^{-t (\bm{T}\bm{\Omega}) }]^{\rm ES}&\text{(B)}
\end{cases}\,.
\end{align}
Assuming $\det(\bm{U}_t)\neq 0$, the subsystem correlation matrix~\eqref{sol_red_correl_matrix} is a solution of the Lyapunov differential equation
\begin{equation}\label{lyap}
    \dot{\bm{\Theta}}_t^{\rm S}=-\bm{A}_t\,\bm{\Theta}_t^{\rm S} - \bm{\Theta}_t^{\rm S}\,\bm{A}_t^\dag +\bm{B}_t\,,
\end{equation}
with
\begin{align}
\label{A}    \bm{A}_t&=-\dot{\bm{U}}_t \,\bm{U}_t^{-1}\,,\\
\label{B}    \bm{B}_t&=\dot{\bm{V}}_t+\bm{A}_t\bm{V}_t+\bm{V}_t\bm{A}_t^\dag\,.
\end{align}
Since the Lindblad ansatz~\eqref{Lindblad_ansatz} yields an equation of motion for the subsystem correlation matrix $\bm{\Theta}_t^{\rm S}$ of the same structure~\eqref{lyap}, the matrices $\bm{A}_t$ and $\bm{B}_t$ are related to the effective Hamiltonian $\bm{h}^{\rm eff}_t$ and the dissipation matrix $\bm{\gamma}_t$ by
\begin{align}
\label{A2}\bm{A}_t&:=
\begin{cases}
\I \bm{h}^{\rm eff}_t + \bm{\gamma}_t + \bm{\gamma}_t^{\rm T} &\text{(F)}\\
-\bm{\Omega}\bm{h}^{\rm eff}_t - \I \bm{\Omega} (\bm{\gamma}_t - \bm{\gamma}_t^{\rm T}) &\text{(B)}
\end{cases}\,,\quad\\    
\label{B2}\bm{B}_t&:=
\begin{cases}
-2 (\bm{\gamma}_t - \bm{\gamma}_t^{\rm T})&\text{(F)}\\
-2 \bm{\Omega} (\bm{\gamma}_t + \bm{\gamma}_t^{\rm T}) \bm{\Omega}&\text{(B)}
\end{cases}\,.
\end{align}
Combining Eqs.~(\ref{A},~\ref{B}) and Eqs.~(\ref{A2},~\ref{B2}), we can finally obtain explicit expressions for the effective Hamiltonian $\bm{h}^{\rm eff}_t$ and the dissipation matrix $\bm{\gamma}_t$ in terms of $\bm{U}_t$ and $\bm{V}_t$. 
Interestingly, $\bm{h}^{\rm eff}_t$ does not depend on the initial environment preparation, but only on the Hamiltonian's couplings in Eq.~\eqref{Hamiltonian2}. 
However, the existence of the time-local form $\mathcal{L}_t=\dot\epsilon_t\circ \epsilon_t^{-1}$ requires the inverse matrix $\bm{U}_t^{-1}$ to exist, which is not guaranteed for all quadratic models in general.

In the following sections, we apply this theoretical framework to (a) a bipartition of a Kitaev chain and (b) a Kitaev chain boundary-coupled to a fully connected free-fermion environment (see Fig.~\ref{fig1}). For these examples, we find numerically that $\bm{U}_t$ remains invertible throughout the whole time evolution, therefore identification (\ref{A} - \ref{B2}) can be made. In both cases, the external environment, whose preparation determines the structure of the time-local generator through the matrix $\bm{V}_t$, is thermal at the inverse temperature $\beta$,
\begin{equation}\label{thermal_state}
    \rho^{\rm E}_0 = \frac{\e^{-\beta H^{\rm E}}}{\tr\left(\e^{-\beta H^{\rm E}}\right)}\,,
\end{equation}
where
\begin{equation}\label{Hamiltoniana_ambiente}
    H^{\rm E}=\frac{1}{4}\sum_{ij=2l+1}^{2L} T_{ij}\;y_i y_j=\frac{1}{4}\sum_{ij=1}^{2(L-l)} T^E_{ij}\;y_{i+2l}\, y_{j+2l}\,,
\end{equation}
is the environment Hamiltonian and $\bm{T}^{\rm E}$ is a coefficient matrix with elements $T^{\rm E}_{ij}=T_{i+2l,j+2l}\;\forall i,j\in[1,2(L-l)]$.
Importantly, the state~\eqref{thermal_state} is Gaussian since the Hamiltonian $H^{\rm E}$ is quadratic. For fermionic Gaussian states, the environment correlation matrix $\bm{\Theta}^{\rm E}_0$ and the coefficient matrix $\bm{T}^{\rm E}$ are related by~\cite{surace2022fermionic,zhang2020lattice,fagotti2013reduced,fagotti2010entanglement}
\begin{equation}\label{environment_thermal_corr}
\bm{\Theta}^{\rm E}_0 = \tanh\left(\beta \bm{T}^{\rm E}/2\right)\,.\qquad \text{(F)}
\end{equation}
Eq.~\eqref{environment_thermal_corr} is used to reconstruct the matrix $\bm{V}_t$. Importantly, in the infinite temperature limit, $\bm{\Theta}^{\rm E}_0=0$ and $\bm{V}_t=0$.

%%%%%%%%%%%%%%%%%%%%%%%%%%%%%%%%%%%%%%%%%%%
%%%%%%%%%%%%%%%%%%%%%%%%%%%%%%%%%%%%%%%%%%%

\section{Subsystem dynamics of a Kitaev chain}
\label{Sec.V}

In this section, we focus on an $L$-site Kitaev chain under periodic boundary conditions,
\begin{equation}\label{Kitaev}
    H=\sum_{i=1}^{L} (c_i^\dag - c_i)\,(c_{i+1}^\dag + c_{i+1})+2g\sum_{i=1}^{L} n_i\,,
\end{equation}
where the $c$'s are fermion operators, $n_i=c_i^\dag c_i$ is the local number operator and $g$ is the chemical potential parameter. For the periodic boundary conditions, $c_{L+1} = c_1$. Again, we consider a bipartition of the full Kitaev chain, consisting of a $l$-site subsystem and a $(L-l)$-site environment, and we aim to extract the generator of the subsystem dynamics. The Hamiltonian~\eqref{Kitaev} is in the form~\eqref{Hamiltonian}, and our theoretical framework can be applied. 
 
For any $l$-site subsystem, one can easily obtain the explicit form of $\bm{U}_t$, Eq.~\eqref{U_t}, in the thermodynamic limit ($L\to\infty$, $l/L\to 0$), thus eliminating any residual finite-size effect (see App.~\ref{App_ExplicitU} for details). However, the matrix elements of $\bm{U}^{-1}_t$ cannot generally be expressed in a simple analytical form. 
For this reason, unless explicitly stated otherwise, the results presented in this section are obtained numerically at relatively low cost, as the evolution is Gaussian preserving. 
Importantly, we numerically find $\bm{U}_t$ to be invertible if $g\neq 0$ and the environment is chosen to be large enough to avoid re-entrant finite-size effects in the time window in which the reduced dynamics is studied.  

\subsection{Single-site subsystem}
\label{Sec.Va}

We start by presenting the results for $l=1$, where both $\bm{h}^{\rm eff}_t$ and $\bm{\gamma}_t$ are $2\times 2$ matrices. Since $\bm{h}^{\rm eff}_t$ is Hermitian and skew-symmetric, its evolution is fully captured by a single real-valued function $\alpha(t)$, namely 
\begin{equation}\label{effect_Ham_1qubit}
    H^{\rm eff}_t=\frac{\I}{2}\alpha(t) y_1 y_2\,.
\end{equation}
On the other hand, two real and non-degenerate eigenvalues are expected for $\bm{\gamma}_t$. We denote them $\Gamma_1(t)$ and $\Gamma_2(t)$.

\subsubsection{Study of the degree of non-Markovianity}
In this section, we study the degree of non-Markovianity of single-site maps as a function of the environment inverse temperature $\beta$ and the chemical potential $g$, using two inequivalent measures of non-Markovianity, namely the Rivas-Huelga-Plenio (RHP)~\cite{rivas2009entanglement} and the Breuer-Laine-Piilo (BLP)~\cite{breuer2009measure,laine2010measure} measures.

As a quantifier of CP-divisibility breaking, the RHP measure relies on estimating the violation of positivity of the Choi matrix~\cite{rivas2009entanglement}, which is equivalent to quantifying the negativity of the couplings $\Gamma_i(t)$~\cite{hall2014canonical}, 
\begin{equation}
    N_{\rm RHP}(t)=\sum_{i=1}^{\mathcal{D}_{\rm S}^2-1}\int_{\substack{\tau\in[0,t]\\ \Gamma_i(\tau)<0}} \diff \tau \,\abs{\Gamma_i(\tau)}\,.
\end{equation}
Notice that, if $\epsilon_t$ is CP-divisible then $\Gamma_i(t)\geq 0$ $\forall i$~\cite{breuer2016colloquium}, which makes $N_{\rm RHP}(t)=0$ as expected. 

On the other hand, the BLP measure relies on the distinguishability of quantum states. Given a pair of arbitrary subsystem states $\rho^{\rm S}_0$ and $\xi^{\rm S}_0$, their trace distance is defined as
\begin{equation}
    d(\rho^{\rm S}_0,\xi^{\rm S}_0):=\frac{1}{2}\| \rho^{\rm S}_0 - \xi^{\rm S}_0\|_{1}=\frac{1}{2}\tr\abs{\rho^{\rm S}_0-\xi^{\rm S}_0}\,.
\end{equation}
This distance is equal to $1$ for orthogonal states, $\rho^{\rm S}_0\perp \xi^{\rm S}_0$, and admits a clear physical interpretation in terms of distinguishability: a larger trace distance corresponds to a higher probability of successfully distinguishing the states $\rho^{\rm S}_0$ and $\xi^{\rm S}_0$~\cite{breuer2016colloquium}. One of the most important properties of the trace distance is its contractivity under CP maps, which implies
\begin{equation}\label{contraction}
    d(\rho^{\rm S}_t,\xi^{\rm S}_t) \leq d(\rho^{\rm S}_0,\xi^{\rm S}_0)\,,\qquad \forall t\geq 0\,,
\end{equation}
for $\rho^{\rm S}_t=\epsilon_t[\rho^{\rm S}_0]$, $\xi^{\rm S}_t=\epsilon_t[\xi^{\rm S}_0]$ and any pair of initial states $\rho^{\rm S}_0$ and $\xi^{\rm S}_0$. Eq.~\eqref{contraction} means that the subsystem-environment interaction can only make the two states less distinguishable in time; this decrease of the trace distance is interpreted as a loss of information from the subsystem to the environment. In contrast, the increase of the distance signals a backflow of information from the environment into the subsystem and the presence of memory effects, since information on the states is temporally stored into the environment and later returns to influence the evolution. The BLP measure~\cite{breuer2009measure,laine2010measure} quantifies non-Markovianity by the increase in distinguishability, measured by a positive derivative of the distance, and reads
\begin{equation}\label{BLP_measure}
    N_{\rm BLP}(t)=\max_{\rho^{\rm S }_0, \,\xi^{\rm S}_0}\int_{\substack{\tau\in[0,t]\\ \partial_\tau d(\rho^{\rm S}_\tau,\xi^{\rm S}_\tau)>0}} \diff \tau \;\partial_\tau d(\rho^{\rm S}_\tau,\xi^{\rm S}_\tau)\,.
\end{equation}
The maximization in~\eqref{BLP_measure} runs in principle over all pairs of initial quantum states $\rho^{\rm S}_0$ and $\xi^{\rm S}_0$. However, as proved in Ref.~\cite{wissmann2012optimal}, maximization can be simplified, since the optimal initial state pairs
lie on the boundary of the Bloch hypersphere and are orthogonal, $\rho^{\rm S}_0\perp \xi^{\rm S}_0$. In our numerical results, the maximum ~\eqref{BLP_measure} is estimated by considering a stochastic sample of $10^5$ pairs or orthogonal pure states. In App.~\ref{App_Bloch}, we provide more explicit and computationally tractable expressions of Eq.~\eqref{BLP_measure} for a single-site subsystem, which will be used throughout this work.  

\begin{figure}
    \centering
    \includegraphics[width=\columnwidth]{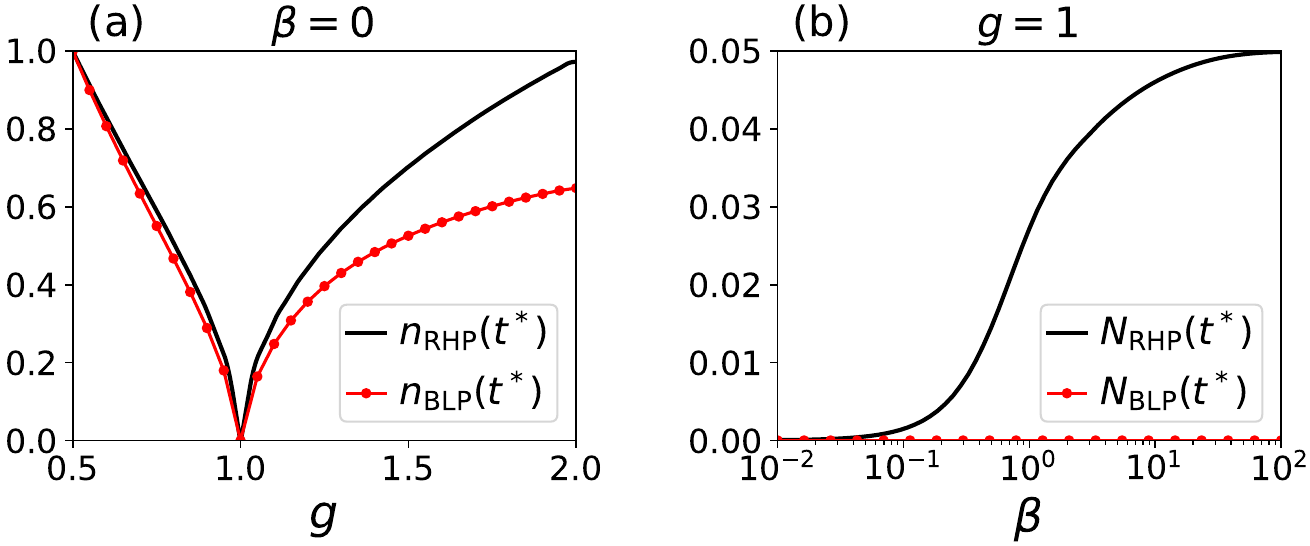}
    \caption{ We plot the non-Markovianity measures as a function of $g$ and $\beta$. Here, we set $l=1$, $L=200$, and time $t^*=25$. In order to compute the maximum in Eq.~\eqref{BLP_measure}, we considered a stochastic sample of $10^5$ pairs of orthogonal pure states. (a) $n_{\rm RHP}(t^*)$ and $n_{\rm BLP}(t^*)$ vs $g\in[0.5,2]$ at $\beta=0$; (b) $N_{\rm RHP}(t^*)$ and $N_{\rm BLP}(t^*)$ vs $\beta\in[10^{-2},10^2]$ at $g=1$ (log-linear scale). }
    \label{fig2}
\end{figure}

In Fig.~\ref{fig2}, we present the RHP and BLP measures as a function of the chemical potential $g$ and inverse temperature $\beta$, at total chain size $L=200$ and time $t^*=25$. Time $t^*=25$ is indeed large enough to be representative of the non-Markovian effects for single-site dynamics, yet small enough to avoid re-entrant boundary effects. We emphasise that, in general, the two measures are inequivalent~\cite{chruscinski2014degree,breuer2016colloquium}, therefore we cannot \emph{a priori} expect the same results. 

In Fig.~\ref{fig2}(a), we set $\beta=0$ and we plot the relative quantities $n_{\rm BLP}(t^*)=N_{\rm BLP}(t^*)/N^{\rm max}_{\rm BLP}(t^*)$ and $n_{\rm RHP}(t^*)=N_{\rm RHP}(t^*)/N^{\rm max}_{\rm RHP}(t^*)$, where $N^{\rm max}_{\rm BLP}(t^*)$ and $N^{\rm max}_{\rm RHP}(t^*)$ denote the maximum values over $g\in[0.5,2]$, such that $n_{\rm BLP}(t^*)$ and $n_{\rm RHP}(t^*)$ lie in the interval $[0,1]$. We observe that the Markovian limit emerges for the critical value $g=1$, with both measures $n_{\rm BLP}(t^*)$ and $n_{\rm RHP}(t^*)$ increasing as we move away from $g=1$.

It is worth noting that, by applying the Jordan-Wigner transformations~\eqref{JW}, the Kitaev chain~\eqref{Kitaev}, when restricted to the odd parity sector, can be mapped onto a transverse field Ising chain. In this mapping, $g$ corresponds to the transverse field. Interestingly, the same emerging Markovian behaviour at the critical value $g=1$ was found in Ref.~\cite{coppola2025learning}. 

In Fig.~\ref{fig2}(b), we plot $N_{\rm RHP}(t^*)$ and $N_{\rm BLP}(t^*)$ as a function of $\beta$ at the critical value $g=1$, which, as discussed above, favors Markovianity. As a result, $N_{\rm RHP}(t^*)$ shows a monotonically increasing behaviour with $\beta$, which implies that large temperatures favour memoryless evolutions. The BLP measure $N_{\rm BLP}(t^*)$ is identically zero for all values of $\beta$ at $g=1$ (for the stochastic sampling considered). 

\subsubsection{Liouvillian}

\begin{figure}
    \centering
    \includegraphics[width=\columnwidth]{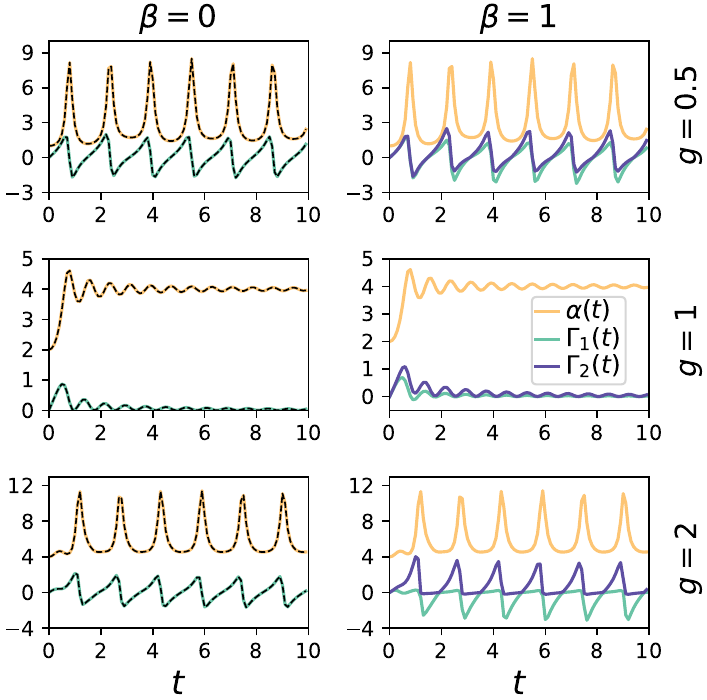}
    \caption{Time evolution of $\alpha(t)$, $\Gamma_1(t)$ and $\Gamma_2(t)$ for a single-site system, $g\in\{0.5,1,2\}$ (top to bottom) and $\beta\in\{0,1\}$ (left to right). The coloured continuos lines represent numerical results for $L=100$, while black dashed lines show analytical results in the thermodynamic limit (see App.~\ref{App_exact-singlesite} for details). }
    \label{fig3}
\end{figure}

In Fig.~\ref{fig3}, we plot $\alpha(t),\,\Gamma_1(t),\,\Gamma_2(t)$ for $g\in\{0.5,1,2\}$ (top to bottom) and $\beta\in\{0,1\}$ (left to right). 
Consistent with analysis in Fig.~\ref{fig2}(a), we find negative rates in some time windows at $g\neq 1$, while for $g=1$ rates are positive at all times for $\beta=0$ and just slightly negative at $\beta=1$ (see Fig.~\ref{fig2}(b)).
At infinite temperature, the spectrum of $\bm{\gamma}_t$ becomes degenerate, i.e., $\Gamma_1(t)=\Gamma_2(t)=\Gamma(t)$. This can be explained by noting that, for $\beta=0$, $\bm{V}_t=0$; consequently, both $\bm{h}^{\rm eff}_t$ and $\bm{\gamma}_t$ are determined solely by $\bm{A}_t$. The generator of the subsystem dynamics is analytically derived (see App.~\ref{App_exact-singlesite} for details), with both $\Gamma(t)$ and $\alpha(t)$ (black dashed lines). At finite temperature ($\beta\neq 0$), this degeneracy is broken. This mechanism will be presented more explicitly in the next section. 

Interestingly, $\Gamma_{1}(t), \Gamma_{2}(t), \alpha(t)$ quickly become periodic for $g\neq 1$, regardless the environment temperature. Consequently, the generator $\mathcal{L}_t$ becomes periodic as well. This observation suggests that the local dynamics can be reconstructed by a Floquet Liouvillian $\mathcal{L}_{\rm F}$, such that the two-time propagator reads
\begin{equation}
    \epsilon_{t+\bar{t},t}=\exp\{\mathcal{L}_{\rm F} \bar{t}\}\,,
\end{equation}
where $\bar{t}$ denotes the period~\cite{prosen2011nonequilibrium,kamleitner2011time,dai2016floquet,iwahori2016long,hartmann2017asymptotic,magazzu2018asymptotic,scopa2019exact}. However, such treatment cannot be extended to larger subsystem sizes $l>1$, as we will show in the multi-site case. 

\subsection{Multi-site subsystem}
In this subsection, we present the results for $l\geq 2$, where both $\bm{h}^{\rm eff}_t$ and $\bm{\gamma}_t$ are $2l\times 2l$ matrices. While matrix $\bm{\gamma}_t$ has $2l$ independent eigenvalues, we find that at the leading order, only four different non-zero eigenvalues are found, two positive and two negative, here denoted by $\Gamma_{1,+}(t),\Gamma_{2,+}(t),\Gamma_{1,-}(t),\Gamma_{2,-}(t)$. This implies that the number of dissipation channels does not scale extensively with the subsystem size. 

\subsubsection{Decay rates and non-Markovianity}

\begin{figure}
    \centering
    \includegraphics[width=1.0\columnwidth]{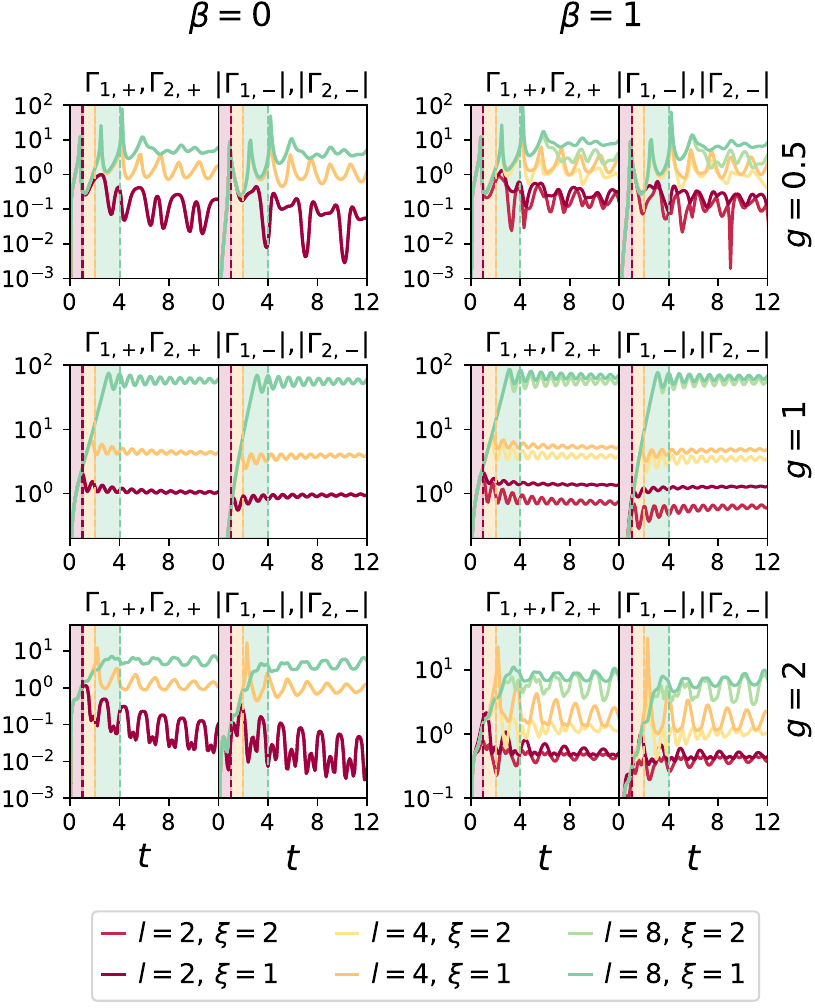}
    \caption{Time evolution of the coupling amplitudes $\{\Gamma_{\xi,\sigma}(t),\,\xi=1,2,\,\sigma=\pm\}$ in linear-log scale, for $l\in\{2,4,8\}$, $L=200$, $g\in\{0.5,1,2\}$ (top to bottom) and $\beta\in\{0,1\}$ (left to right), shown on a semi-log scale. For all cases considered, we observe a transient regime with exponentially growing rates, and an approach to stationarity for $t \gtrsim l/ v_{\max}$. The vertical dashed lines mark the transition between these two regimes. }
    \label{fig4}
\end{figure}

\begin{figure}
    \centering
    \includegraphics[width=1.0\columnwidth]{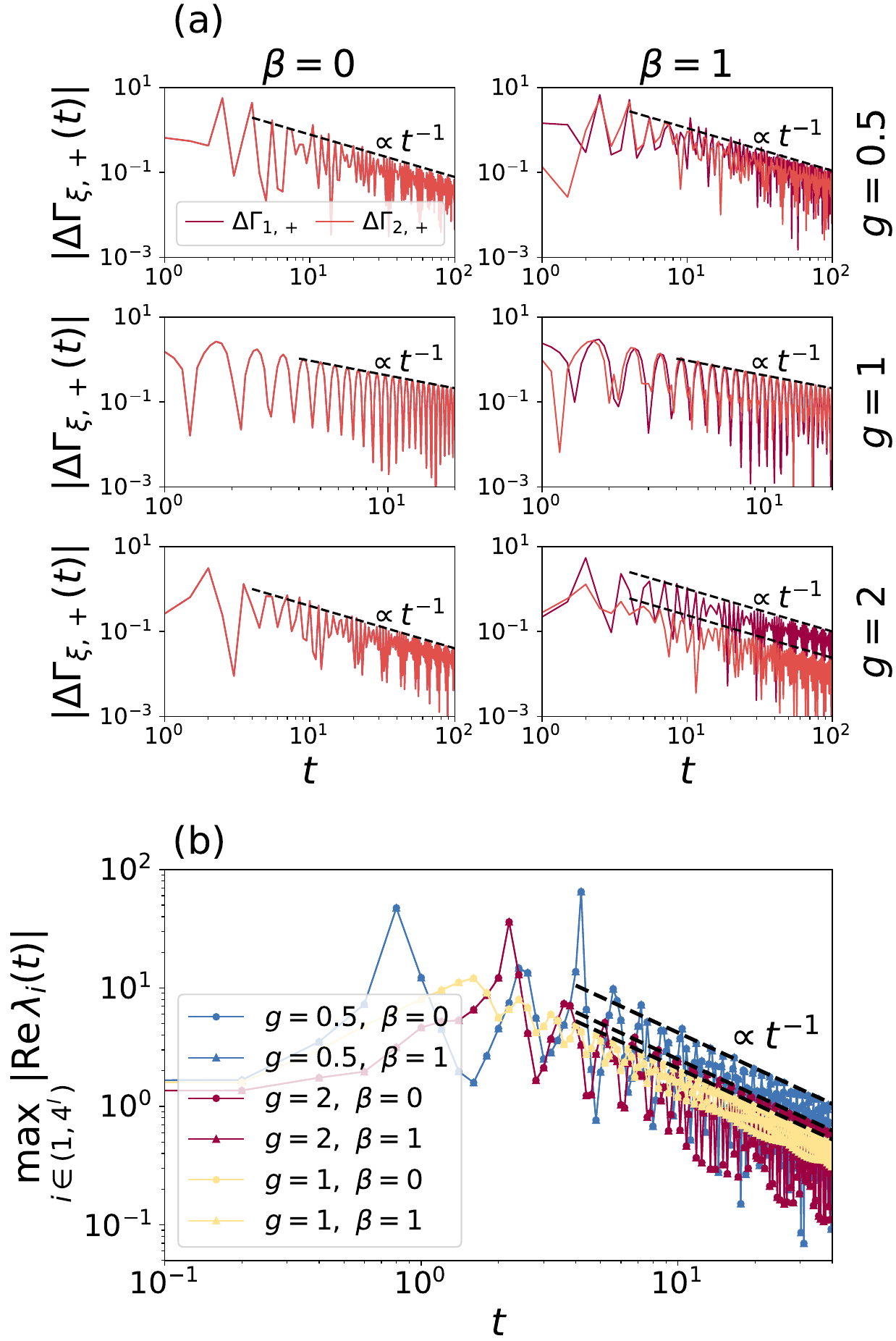}
    \caption{(a) We plot the deviation $|\Delta\Gamma_{\xi,+}(t)|$ of rates $\Gamma_{\xi,+}$ from the stationary value, Eq.~\eqref{eq:DeltaGamma}, as a function of time $t$ in log-log scale, for $l=4$, $\beta\in\{0,1\}$ (left to right) and $g\in\{0.5,1,2\}$ (top to bottom). In all three cases, we observe polynomial decay $t^{-1}$. (b) We plot the dynamics of the real part of Liouvillian's largest eigenvalue $\max_{i\in(1,4^l)}\abs{\Re\lambda_i(t)}$ as a function of time $t$ in log-log scale, for $l=4$, $\beta\in\{0,1\}$ (different markers) and $g\in\{0.5,1,2\}$ (different colours). In all three cases, we again observe polynomial decay $t^{-1}$. }
    \label{fig5}
\end{figure}

In Fig.~\ref{fig4} we plot the time dependence of the four eigenvalues $\{\Gamma_{\xi,\sigma}(t),\,\xi=1,2,\,\sigma=\pm\}$, for $\beta=0$ (left) and $\beta=1$ (right), for three values of the chemical potential $g\in\{0.5,1,2\}$ (top, middle, bottom), and for different subsystem sizes $l\in\{2,4,8\}$. 
Several distinct features can be observed. 

First, for subsystem sizes $l>1$, the evolution is never CP-divisible, i.e. is non-Markovian according to the RHP criterion. Indeed, we always detect two negatives rates $\Gamma_{1,-}(t),\Gamma_{2,-}(t)$ whose amplitude is comparable to the positive ones $\Gamma_{1,+}(t),\Gamma_{2,+}(t)$. 

Second, the evolution of the rates $\Gamma_{\xi,\sigma}(t)$ exhibits two distinct regimes: (i) There is an initial transient regime up to a time $t\sim l/ v_{\max}$, with $v_{\rm max}=2\,{\rm min}(|g|,1)$ denoting the maximum quasiparticle velocity, during which $\abs{\Gamma_{\xi,\sigma}(t)}$ scales exponentially in time. A physical interpretation of this regime will be given in the next section after having discussed the form of associated Lindblad operators. (ii) At times $t \gtrsim l/ v_{\max}$ appears a quasi-stationary regime, in which the  stationary values are approached in a powerlaw fashion, 
\begin{equation}\label{eq:DeltaGamma}
    |\Delta\Gamma_{\xi,\sigma}(t)| \equiv |\Gamma_{\xi,\sigma}(t)-\Gamma_{\xi,\sigma}(\infty)|\propto t^{-1}\,.
\end{equation}
This is shown explicitly in Fig.~\ref{fig5}(a), where we plot the polynomial decay of $|\Delta\Gamma_{\xi,+}(t)|$ as a function of time $t$ in log-log scale, for $l=4$, $\beta\in\{0,1\}$ and $g\in\{0.5,1,2\}$. Similar conclusions can be drawn from the Liouvillian spectrum, $\{\lambda_i(t)\in\mathbb{C},\,i\in(1,4^l)\}$. In particular, the largest absolute real part of the eigenvalues $\lambda_i(t)$, $\max_{i\in(1,4^l)}\abs{\Re\lambda_i(t)}$, decays as $t^{-1}$, reflecting the algebraic long-time relaxation of observables in non-interacting models. This is shown in Fig.~\ref{fig5}(b), where we plot $\max_{i\in(1,4^l)}\abs{\Re\lambda_i(t)}$ as a function of time $t$ in log-log scale, for $l=4$, $\beta\in\{0,1\}$ and $g\in\{0.5,1,2\}$. Here, we should point out that due to negative rates (non-Markovian effects), at $g \neq 1$, the largest non-zero eigenvalue oscillates between positive and negative values. Nevertheless, its algebraic decay towards zero renders physical behaviour of observables.

Third, for the environment initialized at infinite temperature $\beta=0$, the two positive and two negative rates are always degenerate, i.e. $\Gamma_{1,\sigma}(t) = \Gamma_{2,\sigma}(t)$. For an environment initialized at a finite temperature, the degeneracy is present only in the transient regimes and weakly breaks in the quasi-stationary regime. This observation holds also for the single-site case, and we will comment on it below, when discussing the associated Lindblad operators.

\subsubsection{Lindblad operators}
Some of the features observed in the rates $\Gamma_{\xi,\sigma}(t)$ find a simple physical interpretation once we consider the form of the associated dissipator of the subsystem evolution~\eqref{Lindblad_ansatz},
\begin{align}
{\mathcal{D}}[\rho^S_t] 
&= \sum_{ij=1}^{2l}\gamma_{ij}(t)\left[y_i\rho^{\rm S}_t y_j - \frac{1}{2}\{y_jy_i,\rho^{\rm S}_t\}\right] \notag \\
&= \sum_{\substack{\xi=1,2 \\ \sigma=\pm}} \Gamma_{\xi,\sigma}(t) \Big( L_{t}^{\xi,\sigma} \, \rho^{\rm S}_t L_{t}^{\xi,\sigma\;\dagger} - \frac{1}{2}\big \{L_{t}^{\xi,\sigma\;\dagger} L_{t}^{\xi,\sigma}, \rho^{\rm S}_t\big\} \Big)\,,
\end{align}
where $L^{\xi,\sigma}_t$ are the time-dependent Lindblad operators, given by linear combinations of Majorana fermions. Rewriting the Lindblad operators in terms of fermionic creation and annihilation operators, we get
\begin{equation}\label{eq:Lind_weights}
L_{t}^{\xi,\sigma} = \sum_{j=1}^{2l} a^{\xi,\sigma}_j(t) \, c_j^\dagger + b^{\xi,\sigma}_j(t) \, c_j\,. 
\end{equation} 
Each $L_{t}^{\xi,\sigma}$ is a linear superposition of creation and annihilation operators, with weights $a^{\xi,\sigma}_j(t)$ and $b^{\xi,\sigma}_j(t)$ depending on time $t$, the location within the subsystem $j$ and the inverse temperature $\beta$ at which the environment is initialized. 

\begin{figure}
    \centering
    \includegraphics[width=\columnwidth]{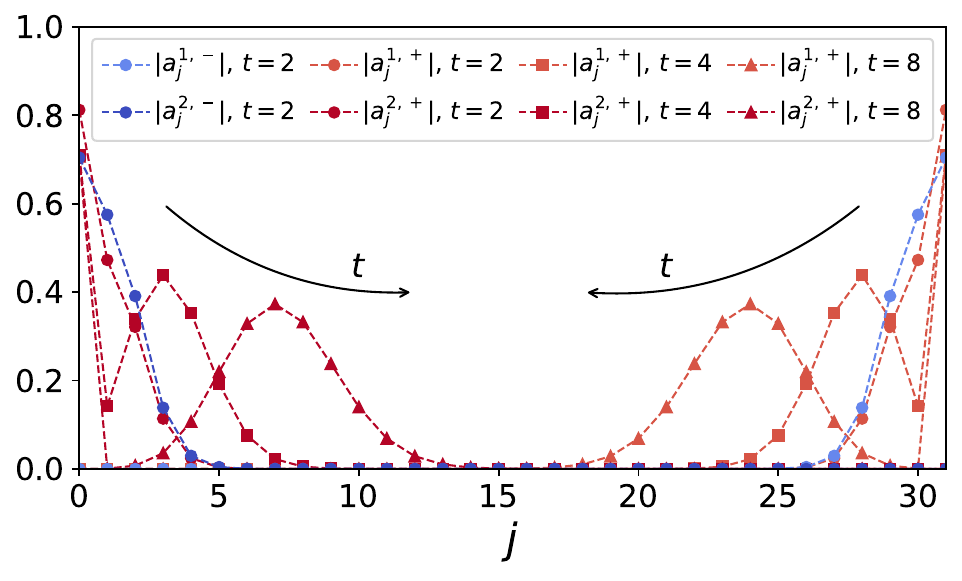}
    \caption{Profiles of the creation weights $|a^{\xi,\sigma}_j|$ at times $t\in\{2,4,8\}$ for $l=32$, $L=400$, $g=1$, $\beta=0$.  The profiles of the Lindblad operators are symmetric, and spread ballistically from the edges to the bulk over time. The annihilation amplitudes $|b ^{\xi,\sigma}_j|$ are omitted as they show the same weights.}
    \label{fig6}
\end{figure}

Following the form of Lindblad operators helps us explain the transient early time regime, during which rates are pairwise degenerate in Fig.~\ref{fig4}. After an appropriate orthogonalization of modes associated with $\{\Gamma_{1,\sigma}(t),\Gamma_{2,\sigma}(t)\}$ at fixed $\sigma$, we obtain two Lindblad operators $\{L^{1,\sigma}_t,L^{2,\sigma}_t\}$ that are localized in the left/right half of the subsystem and spread with time from the point of contact with the environment at the boundaries to the bulk. Fig.~\ref{fig6} shows the profile of creation amplitudes $|a^{\xi,\sigma}_j|$ at different times. In particular, $|a^{+,\sigma}_j|$ are plotted over the full time evolution, while the profiles $|a^{-,\sigma}_j|$ associated with $L_t^{\xi,-}$ are shown only at early times where they differ the most from $|a^{+,\sigma}_j|$. The annihilation amplitudes $|b^{\xi,\sigma}_j|$ are omitted since they behave identically to the creation amplitudes $|a^{\xi,\sigma}_j|$ at $\beta=0$. When analyzing the expansion velocity of the Lindblad profiles to the bulk, we find that the fronts spread at the maximum group velocity $v_{\rm max}=2\,{\rm min}(|g|,1)$. The ballistic nature of quasiparticles within the subsystem and the environment is thus imprinted into a ballistically spreading dissipation, encoded in the time dependence of the support of Lindblad operators. 

\begin{figure}
    \centering
    \includegraphics[width=\columnwidth]{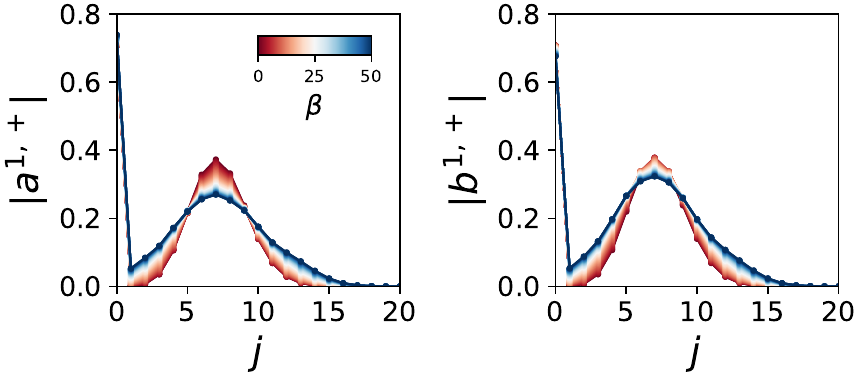}
    \caption{Profiles of the creation and annihilation weights $|a^{1,+}_j|$, $|b^{1,+}_j|$ at times $t=8$ for $l=32$, $L=400$, $g=1$ and $\beta\in[0,50]$. At $\beta >0$, annihilation processes are stronger than creation ones.}
    \label{fig7}
\end{figure}

We recall that the creation and annihilation weights $a^{\xi,\sigma}_j,\, b^{\xi,\sigma}_j$ in Eq.~\eqref{eq:Lind_weights} also depend on the temperature at which the environment is initialized. In Fig.~\ref{fig7} we plot their profiles for different inverse temperatures $\beta\ \in [0,50]$ from red for $\beta=0$ to blue for $\beta=50$. We find that, when the subsystem is in contact with an environment initialized at $\beta=0$, creation and annihilation amplitudes are equal $|a^{\xi,\sigma}_j| = |b^{\xi,\sigma}_j|$, consistent with detailed balance arguments at infinite temperature. As the initial temperature of the environment is lowered, amplitudes at the annihilation operators get larger than the amplitudes at the creation operators. A similar observation is also made if we write the Lindblad operators~\eqref{eq:Lind_weights} in terms of the eigenmodes of the subsystem Hamiltonian $H^{\rm S}$. This is consistent with the interpretation that at $\beta>0$ more energy is taken from the subsystem than is injected into it, while at $\beta=0$ input and output are equal. 

\begin{figure}
    \centering
    \includegraphics[width=\columnwidth]{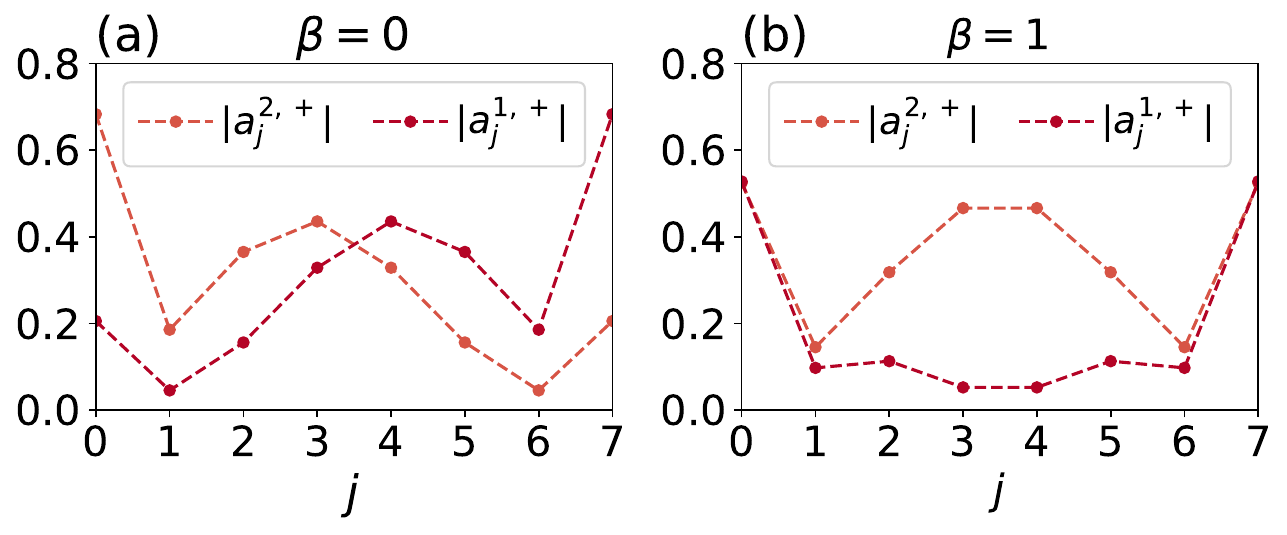}
    \caption{Profiles of the creation weights $|a^{\xi,+}_j|$ in the Lindblad operators at time $t=12$ for $l=8$, $L=400$, $g=1$, $\beta=0$ (left) and $\beta=1$ (right). After the transient, ballistically spreading Lindblad operators have reached half of the subsystem, they hybridize and delocalize over all the $l$ sites. (a) For $\beta=0$, the rates $\Gamma_{\xi,\sigma}(t)$ at $\xi=1,2$ are degenerate and one can choose the corresponding Lindblad operators for $\xi=1,2$ to be symmetric with respect to the middle of the chain. (b) For $\beta >0$, rates $\Gamma_{\xi,\sigma}(t)$ at $\xi=1,2$ are not degenerate and the associated Lindblad operators are distinct.}
    \label{fig8}
\end{figure}

When the fronts of the two symmetric Lindblads collide at time $t= l/ 2 v_{\max}$, they start to hybridize and, for $\beta>0$, the associated rates $\Gamma_{1,\sigma}(t)$ and $\Gamma_{2,\sigma}(t)$ split, lifting the degeneracy as shown in Fig.~\ref{fig4}. After time $t \gtrsim l/ v_{\max}$ and at $\beta>0$, absolute values of creation/annihilation weights are symmetric with respect to the middle of the chain but different for the two Lindblad operators $\xi=1,2$. 

Fig.~\ref{fig8}(a) shows $|a_{j}^{1,+}|$ and $|a_{j}^{2,+}|$ at infinite temperature ($\beta=0$). In this limit, the two Lindblad operators $L^{1,+}_t$ and $L^{2,+}_t$ mirror each other, $|a_{j}^{1,\sigma}| = |a_{L+1-j}^{2,\sigma}|$, $|b_{j}^{1,\sigma}| = |b_{L+1-j}^{2,\sigma}|$, and $\Gamma_{1,+}(t) = \Gamma_{2,+}(t)$. On the other hand, at a finite temperature, this reflection symmetry is broken, $|a_{j}^{1,+}|$ and $|a_{j}^{2,+}|$ no longer mirror each other, and $\Gamma_{1,+}(t)\neq \Gamma_{2,+}(t)$, as shown in Fig.~\ref{fig8}(b).

%%%%%%%%%%%%%%%%%%%%%%%%%%%%%%%%%%%%%%%%%%%
%%%%%%%%%%%%%%%%%%%%%%%%%%%%%%%%%%%%%%%%%%%

\section{Fully connected free-fermion environment}
\label{Sec.VI}

Next, we want to understand how different environment features translate into the generators of subsystem dissipative dynamics, promoting or disfavouring Markovian evolution. For simplicity the environment is prepared at infinite temperature, $\rho^{\rm E}_0=\bm{\mathbb{I}}^{\rm E}/2^{L-l}$, in this section.

For this purpose, we couple the boundaries of a finite Kitaev chain, representing the subsystem, to a fully connected free-fermion environment, Fig.~\ref{fig1}(b). The subsystem Hamiltonian $H^{\rm S}$ is identical to the one considered in Sec.~\ref{Sec.V}. For the interaction Hamiltonian, we assume 
\begin{equation}\label{interaction_Ham}
H^{\rm I} =\I y_1\sum_{j=2l+1}^{2L} p_j y_{j}+\I y_{2l}\sum_{j=2l+1}^{2L} q_j y_{j}\,, 
\end{equation}
where $p_j,\,q_j$ represent the strength of interaction between the subsystem boundary sites and the environmental sites. We collect these weights into corresponding vectors $\bm{p}$ and $\bm{q}$.
Regarding the environment Hamiltonian, we require the environment to exhibit exponential decay of correlations, namely
\begin{equation}\label{eq:bath-corr}
    \langle \e^{\I H^{\rm E} t} y_i \e^{-\I H^{\rm E} t} y_j \rangle \simeq \e^{-\mu t}\delta_{ij}\,,\;\;\text{for}\;\;L\sim \infty\,,
\end{equation}
as a prerequisite of Markovian subsystem evolution in the weak coupling limit $|\bm{p}|,\,|\bm{q}|\ll 1$. Indeed, under these hypotheses, the decoherence time $\tau_{\rm dc}$ is proportional to the inverse of the squared subsystem-environment coupling strength. Therefore, if $\mu\sim \mathcal{O}(1)$, the typical environment correlation time $\mu^{-1}$ is much shorter than $\tau_{\rm dc}$ and the Born-Markov approximation is justified (see App.~\ref{App_weakcoupl}). 

The environment Hamiltonian $H^{\rm E}$, Eq.~\eqref{Hamiltoniana_ambiente}, is stochastically generated in the following way. First, we recall that, for fermions, the matrix $\bm{T}^{\rm E}$ is Hermitian and skew-symmetric. By applying the spectral theorem, $\bm{T}^{\rm E}=\I\bm{Q}\bm{O}\bm{Q}^{\rm T}$, where $\bm{Q}$ is an orthogonal real matrix, $\bm{O}=\bigoplus_{j=1}^{L-l}\begin{bmatrix}
    0 & \nu_j \\
    -\nu_j & 0
\end{bmatrix}$ is the normal block-diagonal form, and $\{\nu_j,-\nu_j\}$ are the real eigenvalues of $\bm{T}_{\rm E}$ occurring in pairs of opposite sign. Inspired by the Caldeira-Leggett model for bosons~\cite{breuer2002theory}, we assume that $\nu_j$ are generated according to the Cauchy distribution $C(\nu_j,\mu)=\frac{1}{\pi}\frac{\mu}{\nu_j^2+\mu^2}$. To generate $\bm{Q}$, we first construct a square matrix $\bm{W}$ with uniformly distributed random entries in $[-1,1]$. Then, we compute its QR decomposition, such that $\bm{W}=\bm{Q}\bm{R}$, where $\bm{Q}$ is an orthogonal matrix and $\bm{R}$ is an upper-triangular matrix. We retain $\bm{Q}$ and discard $\bm{R}$. Such a construction ensures exponential decay of correlations within the environment in the large size limit, Eq.~\eqref{eq:bath-corr}.

Under the hypothesis above, for a single-site subsystem, whose Hamiltonian is $H^{\rm S}=\I g y_1 y_2$, the effective Hamiltonian and the jump rates can be derived analytically in the weak coupling limit (see App.~\ref{App_weakcoupl} for details),  
\begin{align}
\label{H_eff_weak}H^{\rm eff} &= \I \left( g + \frac{2g}{\mu^2+4g^2} (|\bm{p}|^2 + |\bm{q}|^2) \right)y_1 y_2\,,\\
\label{gamma1_weak}\gamma_{1,1} & = \frac{2}{\mu^2+4g^2}\left(\mu |\bm{p}|^2 + 2g \sqrt{|\bm{p}|\,|\bm{q}|}\cos\vartheta\right)\,,\\
\label{gamma2_weak}\gamma_{2,2} & = \frac{2}{\mu^2+4g^2}\left(\mu |\bm{q}|^2 - 2g \sqrt{|\bm{p}|\,|\bm{q}|}\cos\vartheta\right)\,,\\
\label{gamma3_weak}\gamma_{1,2} &= \frac{2}{\mu^2+4g^2}\left(g(|\bm{q}|^2-|\bm{p}|^2)+\mu \sqrt{|\bm{p}|\,|\bm{q}|}\cos\vartheta\right)\,,
\end{align}
where $\cos\vartheta=\bm{p}^{\rm T}\bm{q}/\sqrt{|\bm{p}|\,|\bm{q}|}$. For $|\bm{p}|=|\bm{q}|$ and $\vartheta=\pi/2$, the matrix $\bm\gamma= 2\pi|\bm{p}|^2\, C(2g,\mu)\,\bm{\mathbb{I}}_2$ is diagonal and is proportional to the Cauchy distribution function.

\begin{figure}[t!]
    \centering
    \includegraphics[width=\columnwidth]{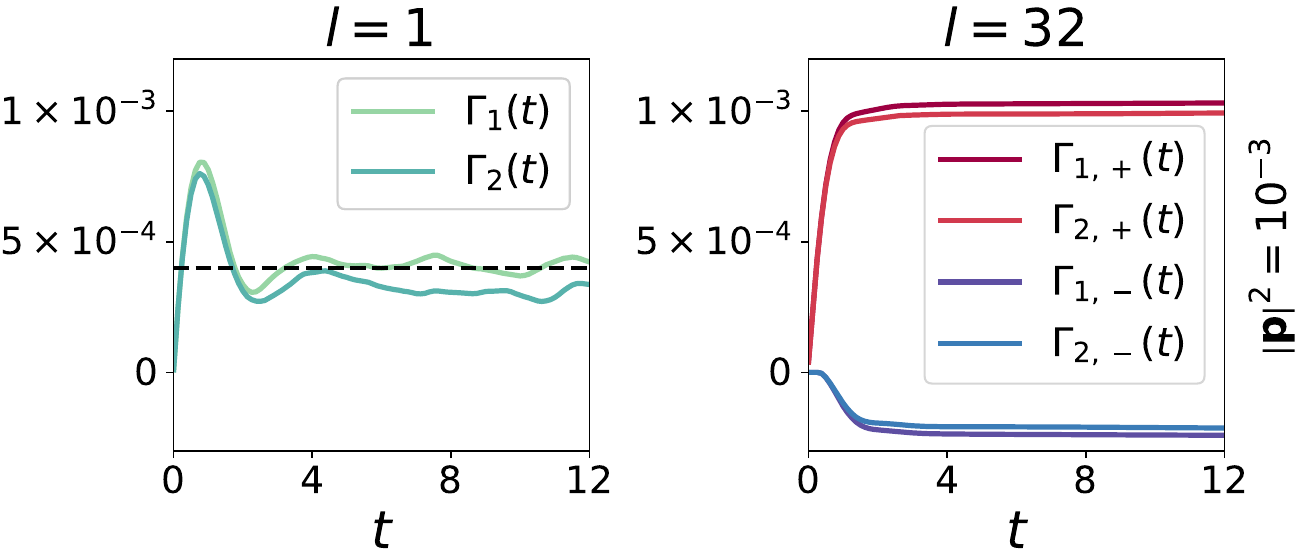}
    \caption{Couplings $\Gamma_j(t)$ (different colours) for $g=1$, $\mu = 1$, $l\in\{1,32\}$ (left to right), $L=2000$, $\vartheta=\pi/2$, and $|\bm{p}|^2=|\bm{q}|^2=10^{-3}$. The black dashed line provides the theoretical results for a single-site subsystem under the weak-coupling approximation, Eqs.~(\ref{gamma1_weak},~\ref{gamma2_weak},~\ref{gamma3_weak}). Due to the weak coupling and exponentially decaying environment correlations, Markovian ($l=1$) or close to Markovian ($l=32$) behaviour is observed.}
    \label{fig9}
\end{figure}

In Fig.~\ref{fig9}, we show the jump rates $\Gamma_j(t)$ as a function of time $t$, for $g=1$, $\mu = 1$, $l\in\{1,32\}$ (left to right), $L=2000$, and $\vartheta=\pi/2$, $|\bm{p}|^2=|\bm{q}|^2=10^{-3}$. Here we set $p_{2L}= -q_{2l+1}$, while all other components are set to zero. Therefore, apart from the amplitudes, the functional form of the interaction Hamiltonian is identical to the one considered in Sec.~\ref{Sec.V}. For the single-site subsystem $l=1$, we find two positive channels with rates that agree with the theoretical prediction, Eqs.~(\ref{gamma1_weak},~\ref{gamma2_weak},~\ref{gamma3_weak}) (black dashed line). For $l>1$, we find again two positive and two negative channels at leading order, denoted by $\Gamma_{1,+}(t),\Gamma_{2,+}(t),\Gamma_{1,-}(t),\Gamma_{2,-}(t)$. Under weak coupling, the two positive rates dominate, indicating proximity to the Markovian limit. As the coupling strengths $|\boldsymbol{p}|, |\boldsymbol{q}|$ increase, the amplitudes of the negative channels grow, signalling the re-emergence of non-Markovian effects.

\begin{figure}[t!]
    \centering
    \includegraphics[width=\columnwidth]{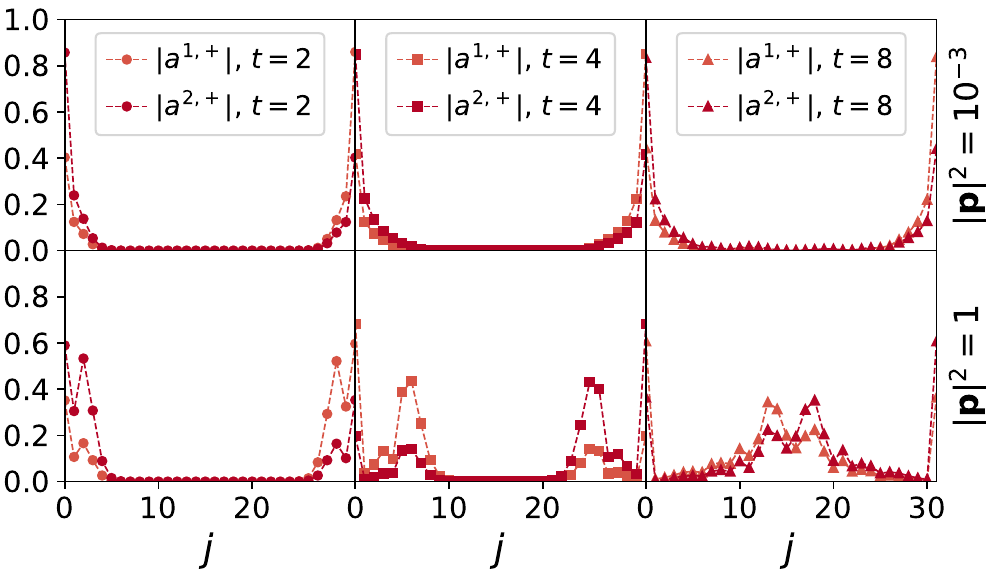}
    \caption{Profiles of the creation weights $|a^{\xi,+}_j|$ in the Lindblad operators at time $t\in\{2,4,8\}$ (left to right) for $l=32$, $L=2000$, $g=1$, $\beta=0$, and $|\bm{p}|^2=|\bm{q}|^2\in\{10^{-3},1\}$ (top to bottom). Lindblad operators remain localized at the edges in the weak-coupling limit, while they feature a ballistic spreading in the support at strong coupling. }
    \label{fig10}
\end{figure}

In Fig.~\ref{fig10}, we show the typical evolution of the weights $|a^{\xi,+}_j|$ of the Lindblad operator $L^{\xi,+}_t$, corresponding to the two positive jump rates $\Gamma_{\xi,+}(t)$. Here, we set $l=32$, $L=2000$, $g=1$, $\beta=0$, and $|\bm{p}|^2=|\bm{q}|^2\in\{10^{-3},1\}$ (top to bottom). 
We clearly see that the profiles are almost perfectly symmetric. However, there is an important distinction: For the weak-coupling that is close to the Markovian regime, the Lindblad operators remain (to zeroth order) localized at the boundaries of the subsystem. On the other hand, in the strong-coupling non-Markovian regime, their weights exhibit the same light-cone evolution as in the example of Sec.~\ref{Sec.V}. 

%%%%%%%%%%%%%%%%%%%%%%%%%%%%%%%%%%%%%%%%%%%
%%%%%%%%%%%%%%%%%%%%%%%%%%%%%%%%%%%%%%%%%%%

\section{Conclusions}
\label{Sec.VII}

In this work, we developed an exact framework for reconstructing the time-local Liouvillians governing the reduced dynamics of subsystems embedded in noninteracting many-body quantum chains. Assuming initially uncorrelated states and Gaussian environments, we showed that the dynamical generator can be reconstructed efficiently from the Lyapunov equation for the subsystem correlation matrix, thereby avoiding the exponential growth in the number of Lindblad operators generally required by the local generator, Eq.~\eqref{general_Lind}.

As a concrete application, in Sec.~\ref{Sec.V} we first considered a Kitaev chain. Partitioning the whole system into the subsystem and the environment, the reduced subsystem dynamics corresponds to a boundary-driven open quantum system, providing an ideal setting to investigate the emergence of effective dissipation from purely unitary many-body evolution. We analyzed both single-site and extended subsystems for initial thermal environments. For single-site subsystems, we implemented the RHP and BLP criteria to study the degree of non-Markovianity, finding out that criticality and large environment temperatures promote Markovianity. 
For extended subsystems, the number of time-dependent Lindblad operators in the Liouvillian is $\mathcal{O}(1)$, reflecting the locality of the subsystem-environment coupling at the subsystem boundary. 
After coupling the subsystem to the environment, the support of Lindblad operators spreads in time from the coupling point to the interior of the subsystem, with the fronts propagating ballistically at the maximum quasiparticle velocity. 
At $t\simeq l/2 v_{\rm max}$ the left- and right-moving fronts collide. Consequently, the previously spatially separated Lindblad operators become delocalized, and the formerly degenerate rates hybridize and split.
By expanding the Lindblad operators in terms of the creation and annihilation fermion operators, we showed that the environment temperature determines the weights of this expansion. While the weights have equal modulus at infinite temperature, finite temperatures introduce an asymmetry which favours the annihilation processes, with their weights getting larger as temperature decreases. As regards the jump rates, at leading order they are paired so that two are positive and two are negative. Interestingly, the two pairs are always degenerate at infinite temperatures, while finite temperatures lift this degeneracy after the early transient regime.

As a second example, in Sec.~\ref{Sec.VI} we coupled the subsystem to a fully connected free-fermion environment, whose spectrum is randomly generated according to a Cauchy distribution, implying an exponential decay of environment correlations. This model was intended to contrast the example in Sec.~\ref{Sec.V} where the polynomial relaxation of the environment correlations disfavours the Markovian regime, and prevents the subsystem dynamics from being interpreted as a sequence of repeated collisions with identically initialized ancillas. When coupling the Kitaev chain to fully connected free-fermion environments, the positive jump rates dominate the subsystem dynamics in the weak-coupling regime, consistently with the Markovian limit, and the Lindblad operators remain localized at the edges of the subsystem. On the other hand, the Lindblad operators become non-local in the strong-coupling regime, once again exhibiting the same light-cone pattern observed in Sec.~\ref{Sec.V}.

Beyond the specific model considered here, our work establishes an exact and computationally efficient route for constructing reduced generators in quadratic many-body systems. It complements approaches treating non-interacting systems weakly coupled to baths, which can be described by generalized scattering theories
\cite{ulcakar24,perfetto23,riggio24,lumia24,marche26}, including cooling protocols \cite{lloyd25,molpeceres25} and Gaussian unravelings \cite{pocklington26}. We expect our results will provide a useful bridge between open quantum systems and nonequilibrium many-body physics, offering new tools to investigate subsystem equilibration, information propagation, and the emergence of effective irreversible dynamics in isolated quantum matter.

%%%%%%%%%%%%%%%%%%%%%%%%%%%%%%%%%%%%%%%%%%%
%%%%%%%%%%%%%%%%%%%%%%%%%%%%%%%%%%%%%%%%%%%

\section*{Acknowledgments}
The authors acknowledge Pavel Orlov, Alessio Lerose and Alexios Christopoulos for fruitful discussions. 
MC and ZL were supported by the QuantERA II JTC 2021 grants QuSiED and T-NiSQ by MVZI, the P1-0044 program of the Slovenian Research Agency and ERC StG 2022 project DrumS, Grant Agreement 101077265. 
MCB was partly supported by the DFG (German Research Foundation) under Germany's Excellence Strategy -- EXC-2111 -- 390814868, and TRR 360 - 492547816, and by the EU-QUANTERA project TNiSQ (BA 6059/1-1).
We also gratefully acknowledge the High Performance Computing Research Infrastructure Eastern Region (HCP RIVR) for funding this research by providing computing resources of the HPC system Vega at the Institute of Information sciences.

%%%%%%%%%%%%%%%%%%%%%%%%%%%%%%%%%%%%%%%%%%%
%%%%%%%%%%%%%%%%%%%%%%%%%%%%%%%%%%%%%%%%%%%

\appendix
\begin{widetext}
\section{Matrix elements of $T$}
\label{App_Tmatrix}

In terms of $\bm{y}$ operators~\eqref{majorana_fermions}, the Hamiltonian~\eqref{Hamiltonian} can be written as
\begin{equation}\label{HamiltonianQ}
H = \frac{1}{4}\bm{y}^\dag \cdot \bm{Q} \cdot\bm{y}\,,    
\end{equation}
where $\bm{Q}$ is a $2L\times 2L$ matrix with elements
\begin{align}
    Q_{2i-1,2j-1}&=K_{ij}+\left(G_{ij}+G_{ji}^*\right)/2\,,\\
    Q_{2i-1,2j}&=\I K_{ij}-\I\left(G_{ij}-G_{ji}^*\right)/2\,,\\
    Q_{2i,2j-1}&=-\I K_{ij}-\I\left(G_{ij}-G_{ji}^*\right)/2\,,\\
    Q_{2i,2j}&=K_{ij}-\left(G_{ij}+G_{ji}^*\right)/2\,,
\end{align}
$\forall i,j\in[1,L]$ and $\bm{G}^{\rm T}=\mp\bm{G}$. The matrix $\bm{Q}$ can be expressed as the sum of the symmetric and the skew-symmetric parts, namely $\bm{Q}=\bm{Q}_{+}+\bm{Q}_{-}$, with $\bm{Q}_{+}=(\bm{Q}+\bm{Q}^{\rm T})/2$ and $\bm{Q}_{-}=(\bm{Q}-\bm{Q}^{\rm T})/2$. Applying the fermionic/bosonic algebra~(\ref{algebraF}, \ref{algebraB}), we get
\begin{align}
\frac{1}{4}\bm{y}^\dag \cdot \bm{Q} \cdot\bm{y}=\begin{cases}
\frac{1}{4}\bm{y}^\dag \cdot \bm{Q}_{-} \cdot\bm{y} + \frac{1}{4}\tr(\bm{Q}_{+}) & \rm{(F)} \\
\frac{1}{4}\bm{y}^\dag \cdot \bm{Q}_{+} \cdot\bm{y} - \frac{1}{4}\tr(\bm{Q}_{-}(\I \bm{\Omega})) & \rm{(B)} 
\end{cases}\,.
\end{align}
We conclude that, up to additive multiples of the identity operator, the Hamiltonian~\eqref{HamiltonianQ} can be written in the form~\eqref{Hamiltonian2}, with $\bm{T}=\bm{Q}_{\mp}$. 

%%%%%%%%%%%%%%%%%%%%%%%%%%%%%%%
%%%%%%%%%%%%%%%%%%%%%%%%%%%%%%%

\section{Matrix elements of $U_t$}
\label{App_ExplicitU}

Diagonalizing the matrix $\bm{T}$ (Eq.~\eqref{Hamiltonian2}) for model \eqref{Kitaev} one may easily derive the elements of the matrix $\bm{U}_t$ (Eq.~\eqref{U_t}) for a subsystem of size $l$ in the thermodynamic limit ($L\to\infty$, $l/L\to 0$), which read
\begin{align}
    U_{2\alpha,2\beta}(t) &= U_{2\alpha-1,2\beta-1}(t) = \int_{-\pi}^{+\pi}\frac{\diff p}{2\pi}\cos\left[(\alpha-\beta)p\right]\cos(\epsilon_p t)\,,\\
    U_{2\alpha-1,2\beta}(t) &=\int_{-\pi}^{+\pi}\frac{\diff p}{\pi}\left(g+\cos(p)\right)\cos\left[(\alpha-\beta)p\right]\frac{\sin(\epsilon_p t)}{\epsilon_p}+\int_{-\pi}^{+\pi}\frac{\diff p}{\pi}\sin(p)\sin\left[(\alpha-\beta)p\right]\frac{\sin(\epsilon_p t)}{\epsilon_p}\,,\\
    U_{2\alpha,2\beta-1}(t) &=-\int_{-\pi}^{+\pi}\frac{\diff p}{\pi}\left(g+\cos(p)\right)\cos\left[(\alpha-\beta)p\right]\frac{\sin(\epsilon_p t)}{\epsilon_p}+\int_{-\pi}^{+\pi}\frac{\diff p}{\pi}\sin(p)\sin\left[(\alpha-\beta)p\right]\frac{\sin(\epsilon_p t)}{\epsilon_p}\,,
\end{align}
$\forall \alpha,\beta\in[1,l]$, with $\epsilon_p=2\left[(g+\cos(p))^2+\sin^2(p)\right]^{1/2}$ being the single-particle eigenvalue. Notice that, at the critical value $g=1$, the matrix elements of $\bm{U}_t$ reduce to
\begin{align}
\label{g1_U1}    U_{2\alpha,2\beta}(t) &= U_{2\alpha-1,2\beta-1}(t) = (-1)^{\alpha-\beta} J_{2(\alpha-\beta)}(4t)\,,\\
\label{g1_U2}    U_{2\alpha-1,2\beta}(t)&=(-1)^{\alpha-\beta+1}J_{2(\alpha-\beta)-1}(4t)\,,\\
\label{g1_U3}    U_{2\alpha,2\beta-1}(t)&=(-1)^{\alpha-\beta+1}J_{2(\alpha-\beta)+1}(4t)\,,
\end{align}
where $J_n(t)$ are the Bessel functions of the first kind.

%%%%%%%%%%%%%%%%%%%%%%%%%%%%%%
%%%%%%%%%%%%%%%%%%%%%%%%%%%%%%

\section{BLP measure for single-site subsystems}
\label{App_Bloch}

In this appendix, we derive more computationally tractable expressions of Eq.~\eqref{BLP_measure} for a single-site subsystem, which will be used throughout this work.  

First, we recall that a fermionic lattice system can be mapped onto a spin-$\frac{1}{2}$ representation by applying the Jordan-Wigner transformation
\begin{equation}\label{JW}
\sigma^{+}_i=\e^{\I \pi \sum_{j<i} n_j}c^\dag_i\,,\qquad     \sigma^{-}_i=\e^{\I \pi \sum_{j<i} n_j}c_i\,,
\end{equation}
where $n_i=c_i^\dag c_i$, and $\sigma^{\pm}_i = (\sigma^x_i \pm \I \sigma^y_i)/2$ are the spin rising and lowering operators. With this convention, we also get $\sigma^z_i=2n_i-1$. 

Restricting to a single site, the Jordan-Wigner strings become trivial, and we get $y_1=c^\dag+c=\sigma^x$ and $y_2=\I(c^\dag-c)=-\sigma^y$, where the site indices are omitted for simplicity. Therefore, the time-local propagator~\eqref{Lindblad_ansatz} for $l=1$ and the effective Hamiltonian~\eqref{effect_Ham_1qubit}, can be easily written in terms of Pauli operators. Since the density matrix of a single qubit can be expressed as $\rho^{\rm S}_t=\frac{1}{2}(\bm{\mathbb{I}}_2+\bm{r}_t\cdot \bm{\sigma})$, where $\bm{r}_t\in\mathbb{R}^3$ denotes the Bloch vector, the subsystem dynamics can equivalently be described by an equation of motion for the Bloch vector. In particular,
\begin{equation}\label{L_bl}
    \dot{\bm{r}}_t=\bm{L}_t \bm{r}_t+\bm{z}_t\,,
\end{equation}
where 
\begin{align}
\bm{L}_t &= 2\begin{bmatrix}
        -\gamma_{2,2}(t) & -\frac{\alpha(t)}{2}-\gamma^{\rm Re}_{1,2}(t) & 0 \\
        \frac{\alpha(t)}{2}-\gamma^{\rm Re}_{1,2}(t) & -\gamma_{1,1}(t) & 0 \\
        0 & 0 & - \tr(\bm{\gamma}_t)
    \end{bmatrix}    \in\mathbb{R}^{3\times 3}\,,\\
\bm{z}_t &= 4 \gamma^{\rm Im}_{1,2}(t)\begin{bmatrix}
        0 \\
        0 \\
        1
    \end{bmatrix}\in\mathbb{R}^{3}\,,
\end{align}
and $\gamma^{\rm Re}_{1,2}(t)=\mathrm{Re}\left(\gamma_{1,2}(t)\right)$, $\gamma^{\rm Im}_{1,2}(t)=\mathrm{Im}\left(\gamma_{1,2}(t)\right)$. 

Let us consider two different quantum states,
\begin{equation}
   \rho^{\rm S}_t=\frac{1}{2}(\bm{\mathbb{I}}_2+\bm{r}^1_t\cdot \bm{\sigma})\,,\qquad  \xi_t^{\rm S}=\frac{1}{2}(\bm{\mathbb{I}}_2+\bm{r}^2_t\cdot \bm{\sigma})\,,
\end{equation}
identified by two Bloch vectors $\bm{r}^1_t$ and $\bm{r}^2_t$. The evolution equation of $\Delta \bm{r}_t=\bm{r}^1_t-\bm{r}^2_t$ reads 
\begin{equation}
    \Delta\dot{\bm{r}}_t=\bm{L}_t \Delta\bm{r}_t\,,
\end{equation}
which admits as a formal solution
\begin{equation}
    \Delta\bm{r}_t= \bm{D}_t \Delta\bm{r}_0\,, 
\end{equation}
where 
\begin{equation}\label{D_bl}
    \bm{D}_t=\mathcal{T}\exp{\int_0^t \diff s\, \bm{L}_s}\,,
\end{equation}
and $\mathcal{T}$ is the time ordering operator. Eq.~\eqref{BLP_measure} can be simplified by observing that the distance between $\rho^{\rm S}_t$ and $\xi_t^{\rm S}$ can also be expressed as 
\begin{equation}
    d(\rho^{\rm S}_t,\xi^{\rm S}_t)=\frac{1}{2}|\Delta \bm{r}_t|\,.
\end{equation}
Therefore, its time derivative reads
\begin{equation}
    \partial_t d(\rho^{\rm S}_t,\xi^{\rm S}_t)=\frac{1}{2}\frac{\Delta\bm{r}_t\cdot \Delta\dot{\bm{r}}_t}{|\Delta \bm{r}_t|}\,,
\end{equation}
and using Eqs.~(\ref{L_bl},~\ref{D_bl}), we obtain
\begin{equation}\label{D9}
    \partial_t d(\rho^{\rm S}_t,\xi^{\rm S}_t)=\frac{1}{2}\frac{\Delta\bm{r}_0^{\rm T} \bm{D}_t^{\rm T} \bm{L}_{+,\,t} \bm{D}_t \Delta\bm{r}_0}{|\bm{D}_t \Delta\bm{r}_0|}\,,
\end{equation}
where
\begin{equation}
    \bm{L}_{+,\,t} =\frac{1}{2}(\bm{L}_{t}+\bm{L}_{t}^{\rm T})= -2\begin{bmatrix}
        \gamma_{2,2}(t) & \gamma^{\rm Re}_{1,2}(t) & 0 \\
        \gamma^{\rm Re}_{1,2}(t) & \gamma_{1,1}(t) & 0 \\
        0 & 0 & \tr(\bm{\gamma}_t)
    \end{bmatrix}\,.    
\end{equation}
Since the maximum in Eq.~\eqref{BLP_measure} has to be computed over the pairs of initial pure orthogonal states, $\bm{r}^2_0=-\bm{r}^1_0$ and $\bm{r}^1_0$ lies on the surface of the Bloch sphere. We conclude that Eq.~\eqref{BLP_measure} can be rewritten as
\begin{equation}\label{BLP_measure_2}
 N_{\rm BLP}(t)=\max_{\bm{n}\in \mathcal{S}}\int_{\substack{\tau\in[0,t]\\ \bm{n}^{\rm T} \bm{D}_\tau^{\rm T} \bm{L}_{+,\,\tau} \bm{D}_\tau \bm{n}>0}} \diff \tau \,\frac{\bm{n}^{\rm T} \bm{D}_\tau^{\rm T} \bm{L}_{+,\,\tau} \bm{D}_\tau \bm{n}}{|\bm{D}_\tau \bm{n}|},
\end{equation}
where $\mathcal{S}=\{\bm{n}\in\mathbb{R}^{3}\,|\,|\bm{n}|=1\}$ represents the surface of the Bloch sphere. Eq.~\eqref{BLP_measure_2} was implemented to produce Fig.~\ref{fig2}. As already mentioned in the main text, a stochastic sample of $10^5$ vectors $\bm{n}\in \mathcal{S}$ was used to compute the maximum in Eq.~\eqref{BLP_measure_2}. 

\section{Generator for single-site subsystems; analytical results for $\beta=0$}
\label{App_exact-singlesite}

Considering the Kitaev chain introduced in Sec.~\ref{Sec.V}, in this appendix we provide analytical expressions for the local generator of a single-site subsystem coupled to an environment at infinite temperature ($\beta=0$). 

According to Eq.~\eqref{environment_thermal_corr}, $\bm{\Theta}^{\rm E}_0=0$ for $\beta=0$. Consequently, Eq.~\eqref{V_t} also implies $\bm{V}_t=0$. The effective Hamiltonian and the couplings read
\begin{equation}
\bm{h}^{\rm eff}_t= -\frac{\I}{2} \left(\bm{A}_t-\bm{A}_t^\dag\right)\,,\qquad \bm{\gamma}_t=\frac{1}{4}\left(\bm{A}_t+\bm{A}_t^\dag\right)\,.
\end{equation}
Since $\bm{A}_t$ is provided by the matrix elements of $\bm{U}_t$, for single-site subsystems ($l=1$), one obtains
\begin{align}\label{E2a}
\bm{h}^{\rm eff}_t &= \frac{\dot\delta(t)\omega(t) - \dot\omega(t)\delta(t)}{\omega(t)^2+\delta(t)^2}\begin{bmatrix}
    0 & \I \\
    -\I & 0
\end{bmatrix}\,,\\ 
\label{E2b}\bm{\gamma}_t &= -\frac{\dot\omega(t)\omega(t)+\dot\delta(t)\delta(t)}{2(\omega(t)^2+\delta(t)^2)}\begin{bmatrix}
    1 & 0 \\
    0 & 1
\end{bmatrix}\,,
\end{align}
where $\omega(t):=U_{1,1}(t)=U_{2,2}(t),\,\delta(t):=U_{1,2}(t)=-U_{2,1}(t)$. Explicitly,
\begin{align}
\omega(t)&= \int_{-\pi}^{+\pi}\frac{\diff p}{2\pi}\cos(\epsilon_p t)\,,\\
\delta(t)&=\int_{-\pi}^{+\pi}\frac{\diff p}{\pi}\left(g+\cos(p)\right)\frac{\sin(\epsilon_p t)}{\epsilon_p}\,.
\end{align}
Therefore, we can identify
\begin{align}
\alpha(t) &= \frac{\dot\delta(t)\omega(t) - \dot\omega(t)\delta(t)}{\omega(t)^2+\delta(t)^2}\,,\\
\Gamma(t) &= -\frac{\dot\omega(t)\omega(t)+\dot\delta(t)\delta(t)}{2(\omega(t)^2+\delta(t)^2)}\,,
\end{align}
as they were introduced in Sec.~\ref{Sec.Va}. Eq.~\eqref{E2b} explicitly reveals the degeneracy observed in Fig.~\ref{fig3}. We notice that $\alpha(0)=2g$, which implies $H^{\rm eff}_0=H^{\rm S}$, and $\Gamma(0)=0$. Therefore, immediately after the quench, the subsystem Hamiltonian dominates the evolution, and then the interactions with the environment progressively enter the process. For completeness, in the spin operators~\eqref{JW} the full Liouvillian reads 
\begin{equation}\label{Liouvillian_inf}
   \hspace{-0.4cm} \mathcal{L}_t[\rho^{\rm S}_t] = -\I\frac{\alpha(t)}{2}\comm{\sigma^z}{\rho^{\rm S}_t} + \Gamma(t)\left(\sigma^x\rho^{\rm S}_t\sigma^x + \sigma^y\rho^{\rm S}_t\sigma^y -2 \rho^{\rm S}_t\right).
\end{equation}
One can notice that, as expected, the infinite temperature state is a stationary state for the Liouvillian~\eqref{Liouvillian_inf}, and the subsystem reaches thermal equilibrium with the external environment.  
Expanding the state $\rho^{\rm S}_t=\frac{1}{2}(\bm{\mathbb{I}}_2+\bm{r}_t\cdot \bm{\sigma})$, with $\bm{r}_t\in\mathbb{R}^3$ being the Bloch vector, one can show that $\rho^{\rm S}_t$ verifies Eq.~\eqref{L_bl}, with  
\begin{equation}
    \bm{L}_t = \begin{bmatrix}
        -2\Gamma(t) & -\alpha(t) & 0 \\
        \alpha(t) & -2\Gamma(t) & 0 \\
        0 & 0 & -4\Gamma(t)
    \end{bmatrix}\,,\qquad \bm{z}_t = 0\,.
\end{equation}
Since $\comm{\bm{L}_{t_1}}{\bm{L}_{t_2}}=0\;\forall t_1,t_2\geq0$, $\bm{D}_t$ is simply given by
\begin{equation}\label{dyn_map_bloch}
    \bm{D}_t=\exp{\int_0^t \diff s\, \bm{L}_s}\,.
\end{equation}
Eq.~\eqref{dyn_map_bloch} can be further simplified by noting that
\begin{align}
    \alpha(t) &=\partial_t \arctan\left(\frac{\delta(t)}{\omega(t)}\right)\,,\\
    \Gamma(t) &= -\frac{1}{4}\partial_t \ln\left(\omega(t)^2+\delta(t)^2\right)\,.
\end{align}
Since $\omega_0=1$ and $\delta_0=0$, we get 
\begin{equation}
    \bm{D}_t= \begin{bmatrix}
        \abs{\omega(t)} & -\delta(t) \,\sgn(\omega(t)) & 0 \\
        \delta(t)\, \sgn(\omega(t)) & \abs{\omega(t)} & 0 \\
        0 & 0 & \omega(t)^2+\delta(t)^2
    \end{bmatrix}\,.
\end{equation}
It is worth noting that
\begin{equation}
    \det(\bm{D}_t)=\exp{\int_0^t \diff s\, \tr(\bm{L}_s)}=(\omega(t)^2+\delta(t)^2)^2\,,
\end{equation}
represents the volume of the accessible states in the Bloch sphere, which must be contractive under P-divisible maps. Since $\partial_t \det(\bm{D}_t)=-8\Gamma(t)\det(\bm{D}_t)$, $\det(\bm{D}_t)$ shows a nonmonotonous decreasing behaviour for $\Gamma(t)>0$. Therefore, the condition $\Gamma(t)<0$ would make the dynamical map non-Markovian according to both the Lorenzo-Plastina-Paternostro (LPP) criterion~\cite{lorenzo2013geometrical} and the RHP criterion (see Sec.~\ref{Sec.Va}). Interestingly, for $g=1$, Eqs.~(\ref{g1_U1}, \ref{g1_U2}, \ref{g1_U3}) yield
\begin{align}
    \Gamma(t) &= J_1(4t)\frac{J_0(4t)+J_2(4t)}{J_0(4t)^2+J_1(4t)^2}\,,\\
    \alpha(t) &= 2 \frac{J_0(4t)^2- J_0(4t)J_2(4t) +2J_1(4t)^2}{J_0(4t)^2+J_1(4t)^2}\,.
\end{align}
Notice that, since $\Gamma(t)\geq 0$ for $g=1$ and $\beta=0$, the dynamical map of the single-site subsystem is CP-divisible, therefore Markovian according to both the RHP and LPP criteria (see Fig.~\ref{fig2}).

%%%%%%%%%%%%%%%%%%%%%%%%%%%%
%%%%%%%%%%%%%%%%%%%%%%%%%%%%

\section{Fermion coupled to a fully connected noninteracting fermion environment in the weak-coupling limit}
\label{App_weakcoupl}

Here, we summarize how to derive the results into Eqs.~(\ref{H_eff_weak},~\ref{gamma1_weak},~\ref{gamma2_weak},~\ref{gamma3_weak}). For this purpose, we largely follow the standard route presented in Ref.~\cite{breuer2002theory}. Here, we restrict to single-site subsystems ($l=1$) and infinite-temperature environments, $\rho^{\rm E}_0\sim \bm{\mathbb{I}}^{\rm E}$. The subsystem Hamiltonian is $H^{\rm S}=\I g y_1 y_2$, while the interaction Hamiltonian is given into Eq.~\eqref{interaction_Ham}. Regarding the environment Hamiltonian~\eqref{Hamiltoniana_ambiente}, $\bm{T}_{\rm E}$ is parametrized as described in Sec.~\ref{Sec.VI}.

In the interaction picture, the evolution of the joint subsystem-environment state is provided by
\begin{equation}\label{E1}
    \dot{\tilde\rho}_t = - \I \comm{\tilde H^{\rm I}_t}{{\tilde\rho}_t}\,,
\end{equation}
where ${\tilde\rho}_t=\e^{\I H^0 t}{\rho_t}\e^{-\I H^0 t}$, $\tilde H^{\rm I}_t=\e^{\I H^0 t}H^{\rm I}\e^{-\I H^0 t}$, $H^0=H^{\rm S}\otimes \bm{\mathbb{I}}^{\rm E}+\bm{\mathbb{I}}^{\rm S}\otimes H^{\rm E}$. After integration of Eq.~\eqref{E1}, we get
\begin{equation}\label{E2}
\dot{\tilde\rho}^{\rm S}_t = -\I \tr_{\rm E}\left(\comm{\tilde H^{\rm I}_t}{\tilde\rho_0}\right) - \int_0^t \diff s \,\tr_{\rm E}\left(\;\comm{\tilde H^{\rm I}_t}{\comm{\tilde H^{\rm I}_s}{\tilde\rho_s}}\;\right)\,.
\end{equation}
To cast Eq.~\eqref{E2} in a closed form, we first observe that the factorized initial state, $\rho_0 = \rho^{\rm S}_0 \otimes \rho^{\rm E}_0$, together with the maximally entangled state $\rho^{\rm E}_0=\bm{\mathbb{I}}^{\rm E}/2^{L-1}$, implies that $\tr_{\rm E}\left(\comm{\tilde H^{\rm I}_t}{\tilde\rho_0}\right)=0$. Moreover, in the weak-coupling limit, the combined subsystem-environment state remains approximately factorized, i.e. $\tilde\rho_t\simeq \tilde\rho^{\rm S}_t \otimes \tilde\rho^{\rm E}_0$.
Therefore, we get
\begin{equation}\label{E3}
\dot{\tilde\rho}^{\rm S}_t \simeq - \int_0^t \diff s \,\tr_{\rm E}\left(\;\comm{\tilde H^{\rm I}_t}{\comm{\tilde H^{\rm I}_s}{\tilde\rho^{\rm S}_s\otimes \tilde\rho^{\rm E}_0}}\;\right)\,.    
\end{equation}
After simple algebraic manipulations, Eq.~\eqref{E3} becomes
\begin{align}\label{E4}
&\dot{\tilde\rho}^{\rm S}_t \simeq  - \int_0^t \diff s \,\left[\langle \bm{p}^{\rm T} \tilde{\bm{y}}_s \bm{p}^{\rm T} \tilde{\bm{y}}_0\rangle_{\rm E} \left( \tilde y_{1,\,t} \tilde y_{1,\,t-s} \tilde\rho^{\rm S}_{t-s} - \tilde y_{1,\,t-s}\tilde\rho^{\rm S}_{t-s}\tilde y_{1,\,t}\right) +\text{h.c.}\right]- \int_0^t \diff s \,\left[\langle \bm{p}^{\rm T} \tilde{\bm{y}}_s \bm{q}^{\rm T} \tilde{\bm{y}}_0\rangle_{\rm E} \left( \tilde y_{1,\,t} \tilde y_{2,\,t-s} \tilde\rho^{\rm S}_{t-s} - \tilde y_{2,\,t-s}\tilde\rho^{\rm S}_{t-s}\tilde y_{1,\,t}\right)+\text{h.c.}\right]\nonumber\\
&\hspace{-0.3cm}- \int_0^t \diff s \,\left[\langle \bm{q}^{\rm T} \tilde{\bm{y}}_s \bm{p}^{\rm T} \tilde{\bm{y}}_0\rangle_{\rm E} \left( \tilde y_{2,\,t} \tilde y_{1,\,t-s} \tilde\rho^{\rm S}_{t-s} - \tilde y_{1,\,t-s}\tilde\rho^{\rm S}_{t-s}\tilde y_{2,\,t}\right)+\text{h.c.}\right]- \int_0^t \diff s \,\left[\langle \bm{q}^{\rm T} \tilde{\bm{y}}_s \bm{q}^{\rm T} \tilde{\bm{y}}_0\rangle_{\rm E} \left( \tilde y_{2,\,t} \tilde y_{2,\,t-s} \tilde\rho^{\rm S}_{t-s} - \tilde y_{2,\,t-s}\tilde\rho^{\rm S}_{t-s}\tilde y_{2,\,t}\right)+\text{h.c.}\right]\,,
\end{align}
where $\tilde{\bm{y}}_s=\e^{\I H^0 s}{\bm{y}}\e^{-\I H^0 s}$ are the Majorana fermions in the interaction picture and $\langle \bullet \rangle_{\rm E}$ represents the average over $\rho^{\rm E}_0$. In the interaction picture, the environmental Majorana operators read 
\begin{equation}\label{E5}
\tilde{y}_{i+2l,\,t}=\sum_{j=1}^{2(L-l)}[\e^{-\I t \bm{T}_{\rm E}}]_{ij} \,y_{j+2l}\,,
\end{equation}
Since $\rho^{\rm E}_0=\bm{\mathbb{I}}^{\rm E}/2^{L-1}$, the environment correlations are
\begin{equation}\label{E6}
\langle y_{i+2l}\,y_{j+2l}\rangle_{\rm E}=\delta_{ij}\,.
\end{equation}
Furthermore, the spectral properties of the matrix $\bm{T}_{\rm E}$ imply that, in thermodynamic limit ($L\to\infty$),
\begin{equation}\label{E7}
[\e^{-\I t \bm{T}_{\rm E}}]_{ij}\overset{L\to\infty}{\longrightarrow}\e^{-\mu t}\,\delta_{ij}\,.
\end{equation}
Combining Eqs.~(\ref{E5},~\ref{E6},~\ref{E7}),  
\begin{align}
   \langle \bm{p}^{\rm T} \tilde{\bm{y}}_s \bm{p}^{\rm T} \tilde{\bm{y}}_0\rangle_{\rm E}&\overset{L\to\infty}{\longrightarrow} \e^{-\mu s} |\bm{p}|^2 \,,\\
   \langle \bm{p}^{\rm T} \tilde{\bm{y}}_s \bm{q}^{\rm T} \tilde{\bm{y}}_0\rangle_{\rm E}=\langle \bm{q}^{\rm T} \tilde{\bm{y}}_s \bm{p}^{\rm T} \tilde{\bm{y}}_0\rangle_{\rm E}&\overset{L\to\infty}{\longrightarrow} \e^{-\mu s} \sqrt{|\bm{p}|\,|\bm{q}|}\cos\vartheta\,,\\
   \langle \bm{q}^{\rm T} \tilde{\bm{y}}_s \bm{q}^{\rm T} \tilde{\bm{y}}_0\rangle_{\rm E}&\overset{L\to\infty}{\longrightarrow} \e^{-\mu s} |\bm{q}|^2\,,
\end{align}
where $\cos\vartheta=\bm{p}^{\rm T}\bm{q}/\sqrt{|\bm{p}|\,|\bm{q}|}$. To cast Eq.~\eqref{E4} in a Lindblad form, we observe that the typical subsystem relaxation time is much longer than the characteristic time scales of the decay of the environmental correlation functions, and we can substitute $\tilde\rho^{\rm S}_{t-s}\simeq \tilde\rho^{\rm S}_{t}$ inside the integral of Eq.~\eqref{E4}. Therefore, Eq.~\eqref{E4} can be rewritten as
\begin{align}\label{E11}
&\dot{\tilde\rho}^{\rm S}_t \simeq  - |\bm{p}|^2 \int_0^t \diff s \,\left[\e^{-\mu s} \left( \tilde y_{1,\,t} \tilde y_{1,\,t-s} \tilde\rho^{\rm S}_{t} - \tilde y_{1,\,t-s}\tilde\rho^{\rm S}_{t}\tilde y_{1,\,t}\right) +\text{h.c.}\right]- \sqrt{|\bm{p}|\,|\bm{q}|}\cos\vartheta\int_0^t \diff s \,\left[\e^{-\mu s}  \left( \tilde y_{1,\,t} \tilde y_{2,\,t-s} \tilde\rho^{\rm S}_{t} - \tilde y_{2,\,t-s}\tilde\rho^{\rm S}_{t}\tilde y_{1,\,t}\right)+\text{h.c.}\right]\nonumber\\
&- \sqrt{|\bm{p}|\,|\bm{q}|}\cos\vartheta\int_0^t \diff s \,\left[\e^{-\mu s} \left( \tilde y_{2,\,t} \tilde y_{1,\,t-s} \tilde\rho^{\rm S}_{t} - \tilde y_{1,\,t-s}\tilde\rho^{\rm S}_{t}\tilde y_{2,\,t}\right)+\text{h.c.}\right]- |\bm{q}|^2 \int_0^t \diff s \,\left[\e^{-\mu s}  \left( \tilde y_{2,\,t} \tilde y_{2,\,t-s} \tilde\rho^{\rm S}_{t} - \tilde y_{2,\,t-s}\tilde\rho^{\rm S}_{t}\tilde y_{2,\,t}\right)+\text{h.c.}\right]\,.
\end{align}
Since 
\begin{align}
    \tilde y_{1,\,t}&=y_1 \cos(2gt) + y_2 \sin(2gt)\,,\\
    \tilde y_{2,\,t}&=y_2 \cos(2gt) - y_1 \sin(2gt)\,,
\end{align}
after integrating Eq.~\eqref{E11} and back-transforming to the Schrödinger picture, we finally get the master equation~\eqref{Lindblad_ansatz}, with the (time-independent) effective Hamiltonian~\eqref{H_eff_weak} and the (time-independent) couplings~(\ref{gamma1_weak},~\ref{gamma2_weak},~\ref{gamma3_weak}).

\end{widetext}

\bibliography{bibliography}

\end{document}